\documentstyle[emulateapj,danonecolfloat]{article}
\lefthead{Adelberger}

\righthead{Measuring the Radiative Histories of QSOs}

\submitted{Received 2004 February 20; Accepted 2004 May 21}

\begin{document}
\newcommand{\lya}{Lyman~$\alpha$}
\newcommand{\lyb}{Lyman~$\beta$}
\newcommand{\degpoint}{\mbox{$^\circ\mskip-7.0mu.\,$}}
\newcommand{\minpoint}{\mbox{$'\mskip-4.7mu.\mskip0.8mu$}}
\newcommand{\secpoint}{\mbox{$''\mskip-7.6mu.\,$}}
\newcommand{\sqdeg}{\mbox{${\rm deg}^2$}}
\newcommand{\squig}{\sim\!\!}
\newcommand{\subsun}{\mbox{$_{\twelvesy\odot}$}}
\newcommand{\et}{{\it et al.}~}
\newcommand{\Rs}{{\cal R}}

\def\ltsima{$\; \buildrel < \over \sim \;$}
\def\simlt{\lower.5ex\hbox{\ltsima}}
\def\gtsima{$\; \buildrel > \over \sim \;$}
\def\simgt{\lower.5ex\hbox{\gtsima}}
\def\propsima{$\; \buildrel \propto \over \sim \;$}
\def\simprop{\lower.5ex\hbox{\propsima}}
\def\arcs{$''~$}
\def\arcm{$'~$}

\twocolumn[
\title{MEASURING THE RADIATIVE HISTORIES OF HIGH-REDSHIFT QSOs WITH THE
TRANSVERSE PROXIMITY EFFECT}

\author{\sc Kurt L. Adelberger\altaffilmark{1}}
\affil{Carnegie Observatories, 813 Santa Barbara St., Pasadena, CA 91101}

\altaffiltext{1}{Carnegie Fellow}

\begin{abstract}
Since the photons that stream from QSOs alter the ionization state of
the gas they traverse, any changes to a QSO's luminosity will produce
outward-propagating ionization gradients in the surrounding intergalactic
gas.  This paper shows that at redshift $z\sim 3$ the gradients will alter the gas's Lyman-$\alpha$
absorption opacity enough to produce a detectable signature
in the spectra of faint background galaxies.  By obtaining noisy 
(S:N$\sim 4$) low-resolution ($\sim 7$\AA) spectra of a several dozen background galaxies in a $R\sim 20'$
field surrounding an isotropically radiating
18th magnitude QSO at $z=3$, it should be possible
to detect any order-of-magnitude changes to the QSO's luminosity
over the previous 50--100 Myr and to measure the
time $t_Q$ since the onset of the QSO's current luminous outburst
with an accuracy of $\sim 5$ Myr for $t_Q\simlt 50$ Myr.
Smaller fields-of-view are acceptable for shorter QSO lifetimes.
The major uncertainty, aside from cosmic variance, will be the shape and orientation
of the QSO's ionization cone.  
This can be determined from the data if 
the number of background sources is increased by a factor of a few.
The method will then provide a direct test of unification models for AGN.
\end{abstract}
]

\section{INTRODUCTION}
\label{sec:intro}

The $10^8$-$M_\odot$ black holes that lie at the heart of nearby bulge 
galaxies are believed to have accreted much of their mass 
in their youths when they shone briefly and brightly as QSOs
(e.g., Yu \& Tremaine 2002).  A great deal
could be learned about the physical processes
that produce super-massive black holes if we could observe 
how their brightnesses varied during this time.  
For example, a black hole that radiates at a fraction
$l\equiv L/L_E$ of its Eddington luminosity and accretes mass with
radiative efficiency $\epsilon\equiv L/\dot M c^2$
requires $t_S\sim M_{\rm BH}/\dot M_{\rm BH}\sim 4\times 10^7 l(\epsilon/0.1)$ yrs
to change its mass significantly.  Showing that QSO outbursts have a much
shorter duration would confirm the popular belief
(e.g., Haehnelt \& Kauffmann 2000; Cavaliere \& Vittorini 2000;
Wyithe \& Loeb 2002; Di Matteo et al. 2003)
that super-massive black holes build up their
masses through numerous accretion episodes.   Various theoretical
attempts to explain the observed correlation between a black hole's
mass and its bulge's velocity dispersion 
(e.g., Silk \& Rees 1998; Adams, Graff, \& Richstone 2001; Burkert \& Silk 2001;
Miralda-Escud\'e \& Kollmeier 2004)
each postulate different physical mechanisms that quench the QSO's
luminosity when it reaches a certain level.  
To the extent that these mechanisms operate on different
time scales, measuring the duration of QSO outbursts should help us
distinguish between them.

The challenge is to figure out from a few short nights or decades of observations
how a QSO's brightness changed over the preceding few million years.
Investigators have resorted to a number of indirect schemes.
By making assumptions about the link between QSOs and either local black holes or
massive potential wells at the redshift of observation, 
several authors have deduced 
that individual QSOs are unlikely to have shone for fewer than 1Myr
or more than 100Myr (e.g., Haehnelt, Natarajan, \& Rees 1998; Richstone et al. 1998;
Martini \& Weinberg 2001; Hosokawa 2002).  Unfortunately
the uncertain underlying assumptions do not seem to allow 
a much more precise limit on the typical QSO lifetime.
In any case statistical arguments like these
will never be able to tell whether a QSO's total radiative lifetime
of (say) $10^7$ yr occurred in one contiguous chunk or was instead split
among numerous shorter bursts, though this distinction could be crucial in
developing a physical picture of black-hole formation.

\begin{figure}[htb]
\centerline{\epsfxsize=9cm\epsffile{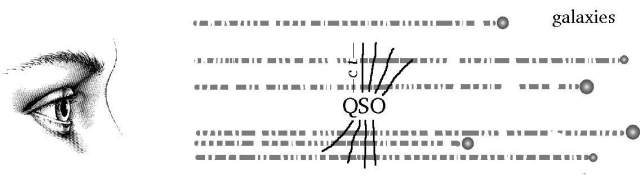}}
\figcaption[f1.eps]{
Schematic view of the underlying concept.  QSOs emit H-ionizing radiation that will destroy some of the
HI in the intergalactic medium out to a maximum radius $ct_Q$, where
$t_Q$ is the time since the onset of QSO activity.  HI in the intergalactic
medium produces absorption lines in the spectra of
background objects.  The absorption is represented by a broken line between
the objects and the observer; breaks in the line mark positions with 
significant HI absorption.  By obtaining spectra of numerous galaxies that lie
behind a high-redshift QSO, one can (a) map the spatial distribution of
intergalactic HI around the QSO, (b) use the shape of the region with lowered HI
absorption to estimate the rough geometry of the intergalactic volume
that has been hit by the QSO's radiation, and (c) deduce the QSO's lifetime
and the level of anisotropy in its emission.
\label{fig:qpemethod}
}
\end{figure}

This paper describes a method that is somewhat less indirect and
that can provide a rough indication
of how any given QSO's luminosity varied over the $50$--100 Myr preceding
the time of observation.  The simple idea behind the method is illustrated
in Figure~\ref{fig:qpemethod}.  
Although we cannot detect the photons a QSO emitted in the past, we can
detect their effect on its surroundings.
As hydrogen-ionizing photons from a QSO propagate
outwards, they destroy neutral hydrogen in the intergalactic medium
and reduce the number of Lyman-$\alpha$ absorption lines in the spectra
of background objects.  The change in the Lyman-$\alpha$ opacity
at radius $r$ should therefore provide some indication of the QSO's luminosity
at the earlier time $t\sim r/c$.  The rest of the paper works out this simple
idea in more detail.
\S~\ref{sec:preliminaries} presents 
a brief review of the relevant intergalactic physics and justifies two assumptions
that will be needed later.  \S\S~\ref{sec:meant} and~\ref{sec:delay}
work out the effect of changes in the QSO's luminosity on
background galaxies' spectra.  
The next two sections discuss
the significance with which the effect can be detected:
\S~\ref{sec:uncertainties} treats the uncertainties from a theoretical
point-of-view and concludes that cosmic variance will be the primary problem,
while \S~\ref{sec:feasibility} discusses the spectroscopic exposure times
that will be necessary on a 10m telescope and 
argues that neither continuum removal nor interstellar absorption lines
will be a major obstacle.
The results from the preceding sections are brought together in one place
in \S~\ref{sec:synthesis}, which presents a sample analysis
of a simulated QSO with a known
radiative history.  Some readers may wish to skip directly to this section.
\S~\ref{sec:binsize} discusses the time resolution that can be achieved 
with this technique.  
My main conclusions are reviewed and criticized in \S~\ref{sec:summary}.

I should state at the outset that this is not the only method 
for constraining the radiative histories of individual QSOs.
Readers may judge for themselves the relative merits of the alternatives
that are listed (e.g.) in the excellent review by Martini (2003).
Nor is the idea behind the method new.  Jakobsen et al. (2003) have used it,
for example,
to estimate an age of $t>10^7$yr for the QSO Q03022-0023 from the lack of absorption
lines in the spectrum of a neighboring QSO.
Schirber, Miralda-Escud\'e, \& McDonald (2004) and Croft (2004)
applied a similar analysis to a larger sample of QSO pairs and
inferred significantly shorter lifetimes. 
What is new, as far as I know, is the 
assumption that the background sources will be numerous faint galaxies
rather than a single bright QSO.
This
complicates the analysis in a number of ways 
but yields a considerably more detailed view of the foreground QSO's radiative history.

\section{PRELIMINARIES}
\label{sec:preliminaries}
The low observed level of Lyman-$\alpha$ absorption from
intergalactic gas at $z\sim 3$ implies that the gas must
be almost completely ionized, with perhaps only one neutral
H atom per million (Gunn \& Peterson 1965).  As a result the typical hydrogen ionizing
photon will travel far before it is absorbed, $\sim 50$ proper 
Mpc ($\Omega_M=0.3$, $\Omega_\Lambda=0.7$, $h=0.65$; Madau, Haardt, \& Rees 1999), 
and one can safely assume that intergalactic
gas is optically thin on length scales significantly smaller.
Radiation from distant sources will permeate this optically thin
gas at a roughly uniform level $J_{\rm bg}$, ionizing residual hydrogen atoms at a rate
$\Gamma^{\rm bg}_\gamma n_{HI}$ where $n_{HI}$ is the neutral hydrogen
density and 
\begin{equation}
\Gamma^{\rm bg}_\gamma\equiv \int_{13.6{\rm eV}}^\infty \!\!\!\!\!\!\!\!\!\!\!dE\, 4\pi J_{\rm bg}(E) \sigma_i(E)/E
\label{eq:gamgam}
\end{equation}
is an integral over energy of the photon number density times the hydrogen 
photoionization
cross-section $\sigma_i$.  
For plausible intensities of the background radiation field, e.g.,
\begin{equation}
J_{\rm bg}(\nu) \sim 10^{-21.3} (h\nu/13.6{\rm eV})^{-1.8} {\rm erg}\,{\rm s}^{-1} {\rm cm}^{-2} {\rm sterad}^{-1} {\rm Hz}^{-1},
\label{eq:jbg}
\end{equation}
and for densities near the cosmic mean,  photon absorption will
be the dominant ionization pathway for hydrogen,
and the neutral fraction $\eta$ of intergalactic gas will adjust itself
until the photoionization rate is equal to the recombination rate
$\alpha_{\mathit HII}n_{\mathit HII}n_e$:
\begin{equation}
\eta\equiv \frac{n_{\mathit HI}}{n_{\mathit HI}+n_{\mathit HII}} \simeq \frac{\alpha_{\mathit HII}n_e}{\Gamma_\gamma}.
\label{eq:neutfrac}
\end{equation}
The recombination coefficient has a temperature dependence that is well fit
by the expression  
$\alpha_{\mathit HII}\simeq 2.11\times 10^{-14} T_6^{-0.7}(1+T_6^{0.7})^{-1}$ cm$^3$ s$^{-1}$
where $T_6$ is the temperature in units of $10^6$ K (Cen 1992).

Now consider what happens to intergalactic
gas when it is hit by a blast of ionizing radiation
from a nearby QSO.  
The photoionization rate increases
by an amount $\Gamma^Q_\gamma$,  given by equation~\ref{eq:jbg}
with the QSO's radiation field $J_Q$ replacing $J_{\rm bg}$,
and the neutral fraction $\eta$ falls to its new equilibrium
value $\eta_{\rm new}=\eta_{\rm old}\Gamma^{\rm bg}_\gamma/(\Gamma^{\rm bg}_\gamma+\Gamma^Q_\gamma)$
on the time scale $|n_{\mathit HI}/\dot n_{\mathit HI}| \sim 1/\Gamma^Q_\gamma$, or
$\sim 10^4$ yr if $J_Q$ is equal to the background
intensity $J_{\rm bg}$.   
After the blast subsides, recombination will raise
the neutral density back to its previous level on the same time-scale.

In contrast to the potentially large swings in the gas's neutral fraction,
any changes to its temperature should be imperceptibly slight.
Although the gas will warm as photo-electrons collide with other
particles and distribute their kinetic energy, the change in its
total thermal energy will be negligible:
photo-electrons are necessarily as rare as neutral atoms ($\sim 1$ppm)
and their typical kinetic energy at ejection 
\begin{eqnarray}
\epsilon_{HI} &=& (\Gamma^Q_\gamma)^{-1}\int_{E_{\rm th}}^{\infty}\!\!\!\!dE\,4\pi J_Q(E)\sigma_i(E)(E-E_{\rm th})/E \\
              &=& E_{\rm th}/(2+s) \simeq 3.6{\rm eV}\quad{\rm for}\quad J_Q(E)\propto E^{-s},\nonumber\\
	      &   & \quad\quad s=1.8,\,\,\sigma_i(E)\propto E^{-3},\,\, E_{\rm th} = 13.6{\rm eV}\nonumber
\label{eq:epsilon_ion}
\end{eqnarray}
is not very
different from the energy per particle in the $\sim 20000$K
undisturbed gas.
The gas received its energy of about an eV per particle
when it was almost completely ionized at earlier times,
and its temperature will hardly be affected by
giving a few eV to the particle per million
that remained neutral.\footnote{I am neglecting the possibility that
the gas around the QSO may be heated by photo-ionization of
atoms other than hydrogen.  In order to cause a significant
temperature change, the product of the photo-ionized atom's number density
$n_X$ and typical ejection energy $\epsilon_X$ must
be comparable to $n_H kT$.  HeII will satisfy this condition before
it is reionized; it is abundant and its photo-electrons' ejection 
energies are large due both to its high ionization potential (see equation~\ref{eq:epsilon_ion})
and to its optical thickness. The latter means that essentially all photons more energetic
than the ionization threshold will be absorbed, not merely the lower energy photons
whose ionization cross-section is greatest (Abel \& Haehnelt 1999).
By redshift $z\sim 3$ the reionization of HeII should be nearly complete,
and my neglect of this heating source is justified.  It would not be if the
redshift were much greater.  }
This shows that the temperature of the gas will not rise appreciably
while its ionization balance adjusts to the increased ionizing background.
Afterwards the gas will be heated by photoionizations at the
same rate as before---the increase in photoionization rate per HI
atom will be exactly compensated by a decrease in the density of HI atoms---and
so the equilibrium temperature will be the same in parts of the IGM that
are and are not illuminated by the QSO's radiation.

Taken together, the results reviewed in this section justify two
assumptions that I will adopt for the remainder of the paper.
(a) If a QSO's luminosity towards solid angle $\Omega$ at time $t$ is
$L(\Omega,t)$, then the intensity of the QSO's radiation field
at position ${\mathbf r}$, $\Omega$ at time $t+|{\mathbf r}|/c$ will
be proportional to $L(\Omega,t)/r^2$.  In other words, intergalactic gas at
larger distances will not be significantly shielded from the QSO's radiation
by intergalactic gas at smaller distances.  (b) The intergalactic temperature 
will not be affected by changes in the intensity of ionizing radiation.
This implies first that a blast of radiation from a QSO 
will alter the ionization balance of intergalactic gas but not its
spatial or velocity structure, and second
that any changes in neutral fraction will be precisely proportional
to the change in the ionizing radiation density (equation~\ref{eq:neutfrac}
with constant $\alpha_{HII}$)
averaged over the last $\sim 10^4$ yr.

\section{MEAN TRANSMISSIVITY VERSUS $J_\nu$}
\label{sec:meant}
We will be able to measure changes in a QSO's ionizing history
with this method only if we can detect spatial changes in the
surrounding density of neutral hydrogen.  Intergalactic HI
density is normally measured by fitting Voigt profiles to
the numerous Lyman-$\alpha$ absorption lines in the
spectra of background QSOs.  This approach is not 
feasible when the background sources are faint galaxies, since
individual Lyman-$\alpha$ forest absorption lines
are hopelessly blended and confused in their noisy low-resolution spectra.
Instead we can only hope to measure $\bar f\equiv \langle e^{-\tau_{{\rm Ly}\alpha}}\rangle$,
the mean transmissivity along a spectral segment 
whose length of a few \AA\ is many times larger than the typical absorption-line spacing
but comparable to the instrumental resolution.
Although a reliance on $\bar f$ is forced on us, $\bar f$
has two advantages over $n_{\rm HI}$ as a probe
of the ionizing background radiation.  First, it receives
significant contributions from the parts of the IGM with middling
HI optical depths $\tau_{{\rm Ly}\alpha}\sim 1$ whose response to
changes in the ionizing radiation are easiest to measure.  In contrast
$\bar n_{\rm HI}$ is dominated by systems with large optical depths,
and since $e^{-\tau}\simeq e^{-2\tau}\simeq 0$ for large $\tau$, 
big changes in the column densities of optically thick systems
can be hard to detect.  Second, the value of $\bar n_{\rm HI}$ along
any particular line-of-sight is strongly affected by whether the line
happens to pierce an especially dense system.  This adds to $\bar n_{\rm HI}$
large random fluctuations that may obscure underlying
changes in the radiation field.  $\bar f$ weights more evenly  
across systems of different column densities and is consequently less
affected by chance fluctuations in the density of matter, a point that
we will elaborate upon in \S~\ref{sec:uncertainties} below.

Let us work out, then, how $\bar f$ responds to changes in the
ionizing background.  
As argued in \S~\ref{sec:preliminaries},
$\tau$ at fixed total hydrogen density is inversely proportional
to the ionizing radiation intensity $J$, and so 
if the radiation intensity changes from $J$ to $bJ$, with $b$
an arbitrary constant, the new mean transmissivity will be
\begin{equation}
\bar f = \int_0^\infty d\tau\, P(\tau) e^{-\tau/b}
\label{eq:barf}
\end{equation}
where $P(\tau)$ is the Lyman-$\alpha$ optical-depth distribution that
existed before the change.

To estimate $P(\tau)$,
I converted published Lyman-$\alpha$ Voigt profile lists for 5 
QSOs\footnote{HS1946+7658 from Kirkman \& Tytler (1997; $z=2.994$)
and Q0636+680, Q0956+122, Q0302-003, and Q0014+813 from Hu et al. 
(1995; $z=3.180$, 3.288, 3.294, 3.366)}
near $z=3$
into lists of the observed Lyman-$\alpha$ optical depth
$\tau$ vs. comoving distance, applied a QSO-dependent
scaling to every $\tau$ to make each QSO's Lyman-$\alpha$ forest
have the same mean transmissivity $\bar f=0.67$ appropriate to
$z=3.00$ (McDonald et al. 2001), and finally constructed a
histogram of the resulting $\tau$ values.

\begin{figure}[htb]
\centerline{\epsfxsize=9cm\epsffile{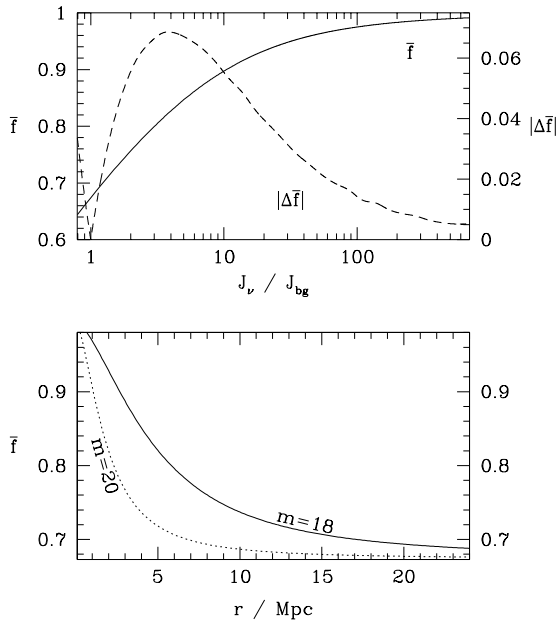}}
\figcaption[f2.eps]{
Top panel: the solid line shows dependence of intergalactic Lyman-$\alpha$ transmissivity 
$\bar f\equiv \langle e^{-\tau_{{\rm Ly}\alpha}}\rangle $ at $z=3$
on the intensity $J_\nu$ of the ionizing radiation field.  The
transmissivity $\bar f\simeq 0.67$ is appropriate to the
actual background ionizing-radiation field $J_{\rm bg}$.  Increases
in the radiation intensity destroy neutral hydrogen and
increase the transmissivity.  The curve shown was calculated
with equation~\ref{eq:barf} from
Voigt-profile fits to the Lyman-$\alpha$ forest at $z\sim 3$ (see text).
The dashed line shows 
$\Delta\bar f\equiv (d\bar f/dJ_\nu)(J_\nu-J_{\rm bg})$, which is roughly 
the change
in $\bar f$ that would result if the perturbation to the
radiation field (i.e., the difference between
$J_\nu$ and $J_{\rm bg}$) were doubled.  
Bottom panel: 
the expected intergalactic transmissivity as a function of distance
from a constantly shining QSO with apparent AB magnitude $m_{912}=18$ 
(solid line) or $m_{912}=20$ (dotted line) at $z=3$.
The QSO's radiation adds to the background radiation $J_{\rm bg}$, which
we are taking to be 
$J_{\rm bg}=5\times 10^{-22}$ erg s$^{-1}$ cm$^{-2}$ Hz$^{-1}$ sr$^{-1}$ 
at $912$\AA, and increases the transmissivity in its
vicinity.  
\label{fig:fluxfigs}
}
\end{figure}

The top panel of figure~\ref{fig:fluxfigs} shows the dependence of
the mean transmissivity on the radiation intensity, calculated numerically
from equation~\ref{eq:barf} using a spline fit to this histogram as an
approximation to $P(\tau)$.
Also shown is $\Delta\bar f\equiv (d\bar f/d\,{\rm ln}J_Q)$,
with $J_Q\equiv J_\nu-J_{\rm bg}$ the QSO's contribution to the
radiation field.  This provides some indication of how 
strongly $\bar f$ responds to large changes in $J_Q$.
The bottom middle panel presents the $J_\nu$--$\bar f$ relationship
in a slightly different way,
as the mean transmissivity as a function of distance $r$ from an isotropically radiating $z=3.0$ QSO
with AB magnitude $m_{912}=18$ or $m_{912}=20$ at rest-frame 912\AA\
that has been radiating forever at a constant rate.
This panel assumes that the background radiation field 
$J_{\rm bg}$ (equation~\ref{eq:jbg})
would have been spatially uniform in the absence of the QSO, which implies
that the actual radiation field is
\begin{equation}
J(r) = \Bigl[1+\Bigl(\frac{r_{\rm eq}}{r}\Bigr)^2\Bigr] J_{\rm bg}
\label{eq:Jr}
\end{equation}
where 
\begin{equation}
r_{\rm eq} \equiv d_L(z) \Bigl[\frac{f_\nu/J_{\rm bg}}{4\pi(1+z)}\Bigr]^{1/2},
\label{eq:req}
\end{equation}
$f_\nu\equiv 10^{-0.4(m_{912}+48.60)}$ is the flux
from the QSO received on earth at wavelength $(1+z)\times$912\AA\
and 
$d_L(z)$ is the QSO's luminosity distance. 

The top panel implies (for example) that if the mean transmissivity in a given
region can be estimated with an uncertainty of $\sigma_f\sim 0.04$,
then a significant change to the QSO's luminosity ($\Delta{\rm ln} J_Q\sim 1$)
will be marginally detectable if the region's distance to the QSO
implies an expected mean transmissivity $0.72\simlt\bar f\simlt 0.93$.
The bottom panel shows that the required distance is $12\simlt r\simlt 2$ proper
Mpc for a QSO with $m_{912}=18$.  Adopting the crude approximation
$t_{\rm delay}\sim -2r/c$ leads to the preliminary guess that the
method should provide reasonable constraints on the QSO's luminosity
for time delays of $-10\simgt t\simgt-80$ Myr.
A more careful treatment is deferred until \S~\ref{sec:synthesis}.

\section{TIME-DELAY SURFACE}
\label{sec:delay}

\begin{figure}[htb]
\centerline{\epsfxsize=9cm\epsffile{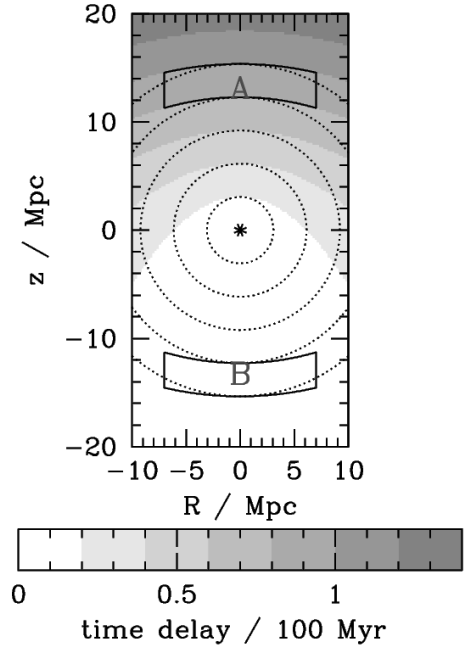}}
\figcaption[f3.eps]{
The time-delay surface.  A QSO at the origin emits hydrogen-ionizing
radiation into the surrounding intergalactic medium.  Concentric circles
show the distance that has been traveled traveled by the photons 
emitted by the QSO at the earlier times $t=-10$, -20, -30, -40, -50 Myr.
Photons from background sources at $z\to\infty$ traverse
this region from top to bottom as they travel to earth at $z\to-\infty$.  
At each point along the way they pass photons that were emitted by
the QSO at a different time in its past.  The shaded contours 
show the emission time of the QSO photons that were illuminating
a given intergalactic region when the observed photons from a background
source passed by.  The strength of the Lyman-$\alpha$ absorption lines 
from this region in the spectrum of the background source will
depend on the QSO's luminosity at the indicated time.  Averaging together
the absorption lines from large regions with the same time delay
(e.g., the region marked $A$) can provide a constraint on the QSO's
luminosity at that time.  Comparing many such regions allows
one to chart historical fluctuations in the QSO's luminosity.
A particularly promising strategy is to estimate the luminosity
at the time corresponding to region $A$ ($\sim -90$ Myr) by
comparing the absorption lines in $A$ to the absorption lines
in a similar region ($B$) on the opposite side of the QSO
that is illuminated by QSO photons emitted at $t\sim 0$.  Most
systematics will cancel out in this binary comparison.  See text.
\label{fig:qpedelay}
}
\end{figure}

The actual situation is slightly more complicated than
Figure~\ref{fig:qpemethod} suggests, since light from the
background sources does not pass the QSO instantaneously.
Instead it encounters material that lies
behind the QSO before it encounters material that lies in front,
and as a result the observed intergalactic absorption
from material behind the QSO will be sensitive to the QSO's luminosity
at an earlier time.
If we define $t=0$ as the time when the QSO emitted the light that
is just now reaching earth,
and if we place the QSO at the origin of
a polar coordinate system where R measures proper displacements along
the plane of the sky and z measures proper displacements in the redshift
direction, then the light from a background galaxy
passes through the intergalactic gas that lies a proper distance 
$z$ behind the QSO at the time $t=-z/c$, and the ionization
balance of the gas at this time is sensitive to the QSO's luminosity
at the earlier time 
\begin{equation}
t_I(R,z) \equiv \bigl[-z-(R^2+z^2)^{1/2}\bigr]/c.
\label{eq:timedelay}
\end{equation}
Figure~\ref{fig:qpedelay} shows the parabolic contours of this
function in astronomically useful units.  
The figure implies, for example,
that the ionization balance of the intergalactic gas that lies
5 Mpc behind the QSO and 2 Mpc to the left will reflect the QSO's luminosity
at time $t=-34$ Myr, while the material that lies directly in front
of the QSO will have an ionization balance that reflects the QSO's 
luminosity at $t=0$ (i.e., its observed luminosity).

Now the field of view of a large optical imagers is around
$40'$, or $\sim 19.9 h_{65}^{-1}$ proper Mpc at $z=3$
($\Omega_M=0.3$, $\Omega_\Lambda=0.7$), and the largest
optical multi-object spectrograph (Dressler, Sutin, \& Bigelow 2003) 
on an 8m-class telescope is not
much smaller.
This makes it relatively easy to obtain spectra of high-redshift
galaxies throughout an $r\sim 20'$ region centered on a QSO at $z=3$,
probing the ionization balance of the IGM in
the region of this plot with $-10 < R < 10$ Mpc.
Because more than 1000 background galaxies with redshift $z\simgt 3.0$ 
and magnitude ${\cal R}\leq 25.5$
will be found in a field of this size with the 
``UV drop-out'' technique (e.g., Steidel et al. 2003),
the Lyman-$\alpha$ absorption lines in these galaxies' spectra can in principle
provide an enormously detailed view of the IGM's ionization balance
near the QSO.  

Assume, then, that we have measured the
intergalactic Lyman-$\alpha$ transmissivity $f\equiv\exp(-\tau_{{\rm Ly}\alpha})$
throughout the region of the figure with $-10<R<10$ Mpc.
A number of ways to estimate the evolution of the QSO's
ionizing luminosity suggest themselves immediately.
One is to divide the region behind the QSO (i.e., the region
with $z>0$ in figure~\ref{fig:qpedelay})
into a number of bins whose edges align with contours of the
time delay surface, then see how the mean transmissivity $\bar f_{\rm obs}$ in each
compares to $\bar f_{\rm exp}$, the expected transmissivity if $L(t)$ were constant.
Region $A$ in figure~\ref{fig:qpedelay} is one such bin.  The problem is estimating $\bar f_{\rm exp}$.
This suggests a slight variation: 
compare $\bar f_{\rm obs}$ in each bin
to $\bar f_{\rm con}$, the mean transmissivity in a bin of the same shape
located on the opposite side of the QSO.  See (e.g.) regions $A$ and $B$ in the figure.
Since $A$ is illuminated by photons emitted at $t\sim -80$ Myr
and $B$ is illuminated by photons emitted at $-10 {\rm Myr}\simlt t< 0$,
the difference in their mean transmissivities will be sensitive
to the any differences in the QSO's luminosity at those two times.
This approach is also not ideal, since 
the mean transmissivity in the control region ($B$) 
is an unnecessarily noisy indicator of the QSO's luminosity
at time $-10 {\rm Myr}\simlt t< 0$; only a small fraction of the volume illuminated
by the QSO's luminosity at $-10 {\rm Myr}\simlt t< 0$ falls in region $B$.
The best approach is probably to find a maximum-likelihood fit of
the data to an appropriate surface.  Since the results of unbinned
maximum-likelihood fits are harder to present in a simple
graphical way, however, I will continue with a paired-bin analysis
for the remainder of this paper but will make one assumption
that should cause the estimated uncertainties to more closely
resemble those of a maximum-likelihood fit: I will
assume that the uncertainties in mean transmissivities of
the control bins are negligibly small.
In other words, I will assume that the uncertainty in the mean
transmissivity in region $B$ of figure~\ref{fig:qpedelay} is 
small compared to the uncertainty for region $A$.  The justification is
that random fluctuations in $B$ can be removed to a large extent
by analyzing the large volume illuminated by the QSO's luminosity 
at $-10 {\rm Myr}\simlt t< 0$.\footnote{
It may help to illustrate this point with a concrete example.
Here is a crude recipe for reducing the noise in the control bins:
fit a low-order polynomial to a plot of the
mean transmissivity in each control bin versus the bin's distance to
the QSO, then use the value of this function in each bin as the control 
transmissivity, rather than the measured transmissivity itself.  
Since we know a priori that 
the mean transmissivity in the control bins ought to be a smoothly
and monotonically declining function of distance to the QSO, deviations
from this behavior must be noise.  They can be largely removed with the 
polynomial fit.  Maximum-likelihood fitting of a surface to the unbinned
data is a more sophisticated implementation of the same idea.}

The fact that so large a region is illuminated by light emitted at $t\sim 0$
is crucial to the success of this approach, since it allows
historical changes in the ionizing luminosity to be measured by comparing
regions that are distributed symmetrically about the QSO.
Various systematics (e.g., the $r^{-2}$ decrease in flux from the QSO, 
bipolar beaming, errors in the continuum fits to the background galaxies'
spectra,
peculiar velocities, gradients in the matter density,  and so on)
should also be symmetric about the QSO, at least in cosmic average,
and so they will cancel out in a front-to-back comparison of 
the HI absorption.  No systematic that I can think of
will have a shape that resembles the time delay
surface in much detail.  
 
The closest candidate might be the gradual expansion of the universe,
which causes the intergalactic material behind the QSO to be slightly
denser than the material in front, changing the
recombination time and producing a slight systematic gradient in the
intergalactic HI density along the line of sight.  This gradient
could mimic
a brightening QSO.
The change
in expansion scale factor as light traverses a region of proper length
$40$ Mpc at $z=3.0$ is $\sim 4$\% ($\Omega_M=0.3$, $\Omega_\Lambda=0.7$, $h=0.65$),
and the corresponding change in mean transmissivity of the IGM
is $\Delta\bar f\sim 0.035$ (e.g., McDonald et al. 2000). 
$\Delta\bar f\sim 0.035$ is comparable to the smallest radiation-related
change we might hope to measure (see below), so this effect cannot
be ignored.  One way to shrink it to insignificance is to divide each
measured transmissivity by $\langle\bar{f}\rangle(z)$, the global mean transmissivity
at its redshift. 
I will assume below that every transmissivity $f$ has been rescaled 
in a way (e.g., divided
by $\langle\bar{f}\rangle(z)$ and then multiplied by $\langle\bar{f}\rangle(3)=0.67$)
that makes this effect negligible.

\section{UNCERTAINTIES}
\label{sec:uncertainties}
Suppose that we have divided the intergalactic volume surrounding the QSO
into different spatial bins, each one illuminated by the light emitted 
by the QSO at a different period $t_i<t<t_i+\Delta t$ in its past,
and that we would like to use the mean Lyman-$\alpha$ transmissivity
$\bar f\equiv\langle e^{-\tau_{{\rm Ly}\alpha}}\rangle$ in each bin
to measure how the QSO's ionizing luminosity has evolved over time.
How reliably can we do this?  \S~\ref{sec:meant} discussed the 
relationship between $\bar f$ and the intensity of the radiation field.
This section is concerned the uncertainty in our estimate of $\bar f$.
The {\it interpretation} of $\bar f$ is subject to further uncertainties,
primarily related to our ignorance of the QSO's true proper distance $z$
and to our assumption that its ionizing radiation is emitted isotropically.
The discussion of these will be deferred until \S\S~\ref{sec:binsize}
and~\ref{sec:summary}.

\subsection{Arithmetic}
\label{sec:arithmetic}
Consider a single spatial bin, for example the volume marked $A$ in 
figure~\ref{fig:qpedelay}, that is being photoionized at the unknown
rate $\Gamma_{\gamma,A}$.  
Let $\bar f_{\rm obs}$ be the mean 
Lyman-$\alpha$ transmissivity we measure along the sight-line segments
that pass through $A$, and let $\bar f_{\rm true}$ be the mean transmissivity
that would be measured in an arbitrarily large intergalactic volume
subjected to the same radiation field.
$\bar f_{\rm obs}$
and $\bar f_{\rm true}$ will differ because (1) our spectra of the background
sources are noisy, (2) we cannot measure the transmissivity throughout
the volume $A$ but must instead rely on an uncertain guess from the
few thinly scattered probes that background sources supply, and (3)
$A$ is not necessarily a fair sample of the universe and could have 
a mean transmissivity that differs from $\bar f_{\rm true}$ for reasons
unrelated to the intensity of ionizing radiation, e.g., if the random
fluctuations from inflation gave it an unusually high or low density 
of matter.

\begin{figure}[htb]
\centerline{\epsfxsize=9cm\epsffile{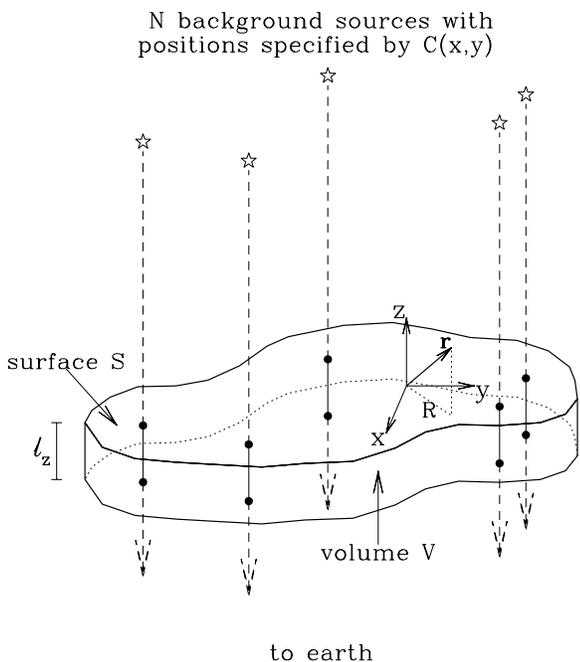}}
\figcaption[f4.eps]{
Nomenclature for \S~\ref{sec:uncertainties}.
The intergalactic volume $V$ is pierced by lines-of-sight to
many background sources.  $V$ is a generalized right cylinder
of height $l_z$ whose cross-section has the arbitrary shape $S$.
Light from the background sources at the top of the figure
moves parallel to the $z$ axis, and normal to $S$, as it makes
its way to towards earth, far below the bottom of the figure.
The $N$ background sources have $(x,y)$ positions that are randomly
distributed within $S$.  
\label{fig:blob}
}
\end{figure}

For simplicity I will obtain estimates of the variance from
(1)--(3) by approximating the actual spatial bin
$A$ as the volume $V$ that lies between two identically shaped
parallel surfaces that are separated by distance $l_z$
(see figure~\ref{fig:blob}).  Generalizing these results to
the true geometry is conceptually trivial.
In the simplified case, the variance due to (1) is simply
\begin{equation}
\sigma^2_{\rm noise} \simeq \frac{1}{N}\Bigl(\frac{\bar f_{\rm true}}{\cal S}\Bigr)^2
\label{eq:s2noise}
\end{equation}
if we have spectra of identical signal-to-noise ratio ${\cal S}$
for each of the $N$ background sources.\footnote{${\cal S}$ should be calculated for wavelength
bins whose size corresponds to the depth $l_z$ of volume $V$, of course.}
The variance due to (2) is related, in a way that is specified below, 
to the variance of the mean transmissivity
among randomly placed sightline segments of length $l_z$,
\begin{equation}
\sigma_l^2 = \Bigl(\frac{1}{2\pi}\Bigr)^3 \int d^3k P({\mathbf k})\Bigl[\frac{\sin(k_zl_z/2)}{k_zl_z/2}\Bigr]^2,
\label{eq:s2line}
\end{equation}
and similarly the variance due to (3) is related to the variance
of the mean transmissivity among randomly placed volumes of
shape $V$,
\begin{equation}
\sigma_V^2 = \Bigl(\frac{1}{2\pi}\Bigr)^3 \int d^3\,r\, P({\mathbf k})\, |W_S(k_x,k_y)|^2\, \Bigl[\frac{\sin(k_zl_z/2)}{k_zl_z/2}\Bigr]^2.
\label{eq:s2volume}
\end{equation}
Here $P({\mathbf k})$ is the power-spectrum of transmissivity
fluctuations
and $W_S(k_x,k_y)$ is the Fourier transform of surface $S$.
Equations~\ref{eq:s2line} and~\ref{eq:s2volume} are both special cases of
Parseval's relationship 
\begin{equation}
\sigma^2 = \Bigl(\frac{1}{2\pi}\Bigr)^3 \int d^3k P({\mathbf k})|W({\mathbf k})|^2
\label{eq:s2general}
\end{equation}
between the power-spectrum and the variance of a random field
that has been
averaged over a volume whose shape has the Fourier transform $W({\mathbf k})$.

The first source of uncertainty, measurement errors, will simply add 
in quadrature to the others.  To understand how the second and third sources
contribute to the total uncertainty, 
consider a given
set $C$ of random galaxy positions $\{x,y\}$, which can be represented as a sum
of Dirac delta functions, $C=(1/N)\sum_j\delta(x-x_j)\delta(y-y_j)$.\footnote{Since
galaxies are spatially extended, $C$ should actually be the sum of functions with
finite width, not the sum of delta functions.  The delta-function
approximation assumes that the bright part of each galaxy is small
compared to the coherence length of intergalactic absorption. This appears to be the case
at $z\sim 3$.}
(The factor of $1/N$ indicates
our decision to average rather than
sum the mean transmissivities from different galaxies.)
Equation~\ref{eq:s2general} implies that the variance of $\bar f$
(i.e., of the transmissivity averaged over each of the $N$ sightline segments
of length $l_z$ that pass through volume $V$)
for this arrangement of background sources is
\begin{equation}
\sigma^2 = \frac{1}{(2\pi)^3}\int d^3k\,P({\mathbf k})|C^k(k_x,k_y)|^2\Bigl[\frac{\sin(k_zl_z/2)}{k_zl_z/2}\Bigr]^2
\label{eq:var_posknown}
\end{equation}
where
\begin{equation}
|C^k({\mathbf k})|^2 = \frac{1}{N^2} \sum_{lm} \exp[i{\mathbf k}\cdot ({\mathbf r}_{l}-{\mathbf r}_{m})]
\label{eq:cki}
\end{equation}
is the powerspectrum of $C$, ${\mathbf r}_m\equiv x_m\hat{\mathbf x} + y_m\hat{\mathbf y}$
is a vector specifying the $x,y$ position of the $m$th background source,
and $\hat{\mathbf x}$ and $\hat{\mathbf y}$ are the usual unit vectors.
Different arrangements of background sources will result in somewhat different
variances, but the average variance among all sets $C$
of random galaxy positions\footnote{in the limit of weak angular clustering} is
given by equation~\ref{eq:var_posknown} with 
$|C^k({\mathbf k})|^2$ replaced by the expectation value
$\langle|C^k({\mathbf k})|^2\rangle$.  The latter can be calculated by
integrating over all random galaxy positions that lie behind the volume $V$:
\begin{eqnarray}
\langle|C^k({\mathbf k})|^2\rangle&\equiv&\frac{1}{N^2} \sum_{lm} \Biggl[\int_S dx_l\,dy_l\,dx_m\,dy_m\,e^{i{\mathbf k}\cdot ({\mathbf r}_{l}-{\mathbf r}_{m})}\nonumber\\
 & & \quad\quad\quad\quad\quad\,\, \Bigg/ \int_S dx_l\,dy_l\,dx_m\,dy_m\Biggr].
\end{eqnarray}
The $N(N-1)$ terms in this sum with $l\neq m$ are each equal to 
$|W_S(k_x,k_y)|^2$, the powerspectrum of $S$,
while the $N$ terms with $l=m$ are each equal to unity.
Substitution into equation~\ref{eq:var_posknown} shows that the expected
variance from the second and third sources of uncertainty is 
a linear combination of the variance in a volume of shape $V$ and
the variance along a line segment of length $l_z$:
$\sigma^2_{2,3} = \frac{N-1}{N}\sigma_V^2 + \frac{1}{N}\sigma_l^2$.
This pleasantly simple result makes intuitive sense.  As $N\to\infty$,
sightlines to the background objects sample almost every part of the
volume $V$.  The mean transmissivity along the sightlines approaches
the mean transmissivity within $V$ arbitrarily closely and $\bar f_{\rm obs}$
becomes as reliable an estimator of $\bar f_{\rm true}$ as
the volume-averaged transmissivity itself.  In the opposite limit, $N=1$,
the volume is pierced by a single skewer.  In this case
the footprint $S$ of the volume (and therefore $\sigma_V$) becomes
irrelevant; the mean transmissivity
along the skewer is determined by physics on the other side of
the universe, not by anything that influences $S$ (e.g.,
our choice of telescope, instrument, pointing, etc.),
and there is no reason that the variance of the transmissivity along
this sightline segment should be any different than the variance
of a similar segment that is randomly placed.

Adding in the variance from measurement errors to
the above expression for $\sigma^2_{2,3}$, we arrive at our final
expression for the total variance of $\bar f$:
\begin{equation}
\sigma^2_{\bar f} = \frac{N-1}{N}\sigma_V^2 + \frac{1}{N}\sigma_l^2 + \frac{1}{N}\Bigl(\frac{\bar f_{\rm true}}{\cal S}\Bigr)^2.
\label{eq:s2fbar}
\end{equation}

It remains to find numerical values for the constants $\sigma^2_l$ and
$\sigma^2_V$. 

\begin{figure}[htb]
\centerline{\epsfxsize=9cm\epsffile{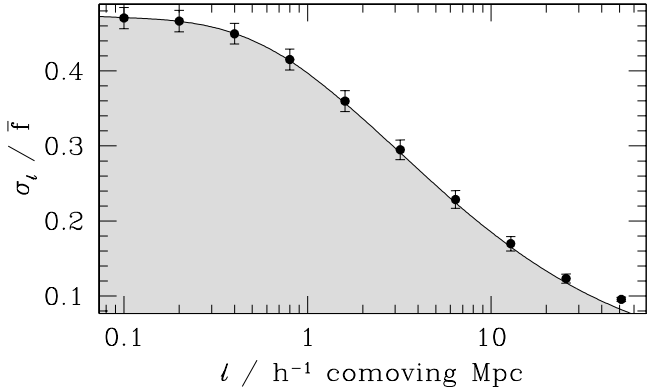}}
\figcaption[f5.eps]{
The r.m.s.~variation of the normalized transmissivity 
$\bar f/\langle \bar f\rangle(z)$ along skewers of different lengths.
Points with error bars are for the 7 primary QSOs from Adelberger et al. (2003);
the shaded curve is the prediction of equations~\ref{eq:pkmcd}
and~\ref{eq:s2line}.  The observed values at the largest $l$ are unreliable
due to uncertainties in the continuum fitting.
\label{fig:s2l_vs_l}
}
\end{figure}

High signal-to-noise QSO spectra can provide a robust empirical estimate of the
variance of the mean transmissivity along a line
segment of length $l_z$.  Figure~\ref{fig:s2l_vs_l} shows
the value I find, as a function of $l_z$, for the primary sample
of 7 QSOs at redshift $z\sim 3$ discussed in Adelberger et al. (2003).

\begin{figure}[htb]
\centerline{\epsfxsize=9cm\epsffile{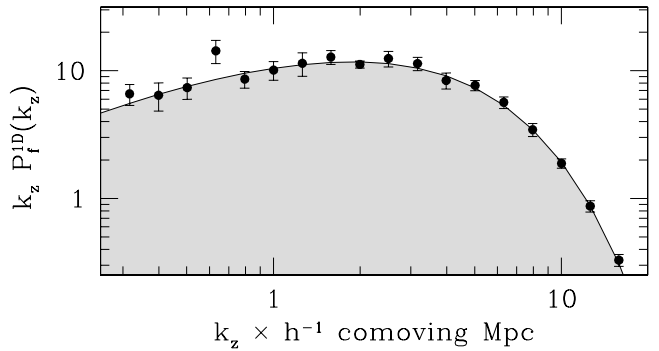}}
\figcaption[f6.eps]{
The one-dimensional power-spectrum of the Lyman-$\alpha$ forest.
Points are from the observations of McDonald et al. (2001);
the shaded curve is the prediction of equation~\ref{eq:pkmcd}.
\label{fig:p1d}
}
\end{figure}

$\sigma_V^2$ is more difficult.  It depends on the three-dimensional
power-spectrum of the Lyman-$\alpha$ forest, which has not been measured
due to a lack of close QSO pairs.  A rough estimate of $\sigma_V^2$
can be obtained by scaling from $\sigma_l^2$, $\sigma_V^2\equiv\beta\sigma_l^2$,
with $\beta$ a constant that can be estimated
by assuming a shape for the powerspectrum and
numerically integrating equations~\ref{eq:s2line} and~\ref{eq:s2volume}.
According to McDonald (2003), 
the transmissivity powerspectra found in numerical simulations 
the high-redshift intergalactic medium have the form
\begin{equation}
P(k_R,k_z) \propto \Bigl[1+\beta k_z^2/k^2\Bigr]^2 P_L(k) D(k_R,k_z)
\label{eq:pkmcd}
\end{equation}
where $k_z$ and $k_R$ specify the wavenumber in polar coordinates,
$k^2\equiv k_z^2 + k_R^2$, $P_L(k)$ is the linear powerspectrum of cold 
dark matter,
\begin{equation}
D(k_R,k_z)\equiv \exp\Bigl[\Bigl(\frac{k}{k_{nl}}\Bigr)^{\alpha_{nl}} - \Bigl(\frac{k}{k_p}\Bigr)^{\alpha_p} - \Bigl(\frac{k_z}{k_v(k)}\Bigr)^{\alpha_v}\Bigr],
\end{equation}
and $k_v(k)\equiv (1+k/k_v')^{\alpha'_v} k_{v0}$.
McDonald (2003) supplies values for the numerical constants in these
equations appropriate to redshift $z\sim 2$.  Adjusting his values
slightly,
to $\beta=1.73$, $k_{nl}=5.2$, $\alpha_{nl}=0.673$, $k_p=8.51$,
$\alpha_p=1.14$, $k_{v0}=0.603$, $k_v'=0.085$, $\alpha'_v=0.432$, $\alpha_v=1.59$,
and adopting a $\Gamma=0.2$ linear powerspectrum with a $P_L(k)\propto k^{0.95}$
long-wavelength limit (Bardeen et al. 1986),
I find a transmissivity powerspectrum that correctly predicts both the observed
dependence of $\sigma^2_l$ on $l_z$ at redshift $z=3$ 
(figure~\ref{fig:s2l_vs_l})
and the observed one-dimensional transmissivity powerspectrum 
$P^{1D}(k_z)\equiv2\pi\int dk_R k_R P(k_R,k_z)$ at redshift $z=3$ 
(figure~\ref{fig:p1d}).  This model powerspectrum is presumably
a reasonable approximation to the true powerspectrum. Integrating it
numerically in
the following special cases of equation~\ref{eq:s2volume},
\begin{eqnarray}
\sigma_V^2&=&\Bigl(\frac{1}{2\pi}\Bigr)^2 \int dk_R\,dk_z k_R P(k_R,k_z)\Bigl\{\Bigl[\frac{\sin(k_zl_z/2)}{k_zl_z/2}\Bigr]^2\times\nonumber\\
& &\quad\quad\quad\quad\quad\quad\quad\quad\Bigl[\frac{2J_1(k_RR)}{k_RR}\Bigr]^2\Bigr\}
\label{eq:s2volume_azi}
\end{eqnarray}
or
\begin{eqnarray}
\sigma_V^2&=&\Bigl(\frac{1}{2\pi}\Bigr)^2 \int dk_R\,dk_z k_R P(k_R,k_z)\Bigl\{\Bigl[\frac{\sin(k_zl_z/2)}{k_zl_z/2}\Bigr]^2\times\nonumber\\
& &\quad\quad\quad\quad \Bigl[\frac{2R_oJ_1(k_RR_o)-2R_iJ_1(k_RR_i}{k_R(R_o^2-R_i^2)}\Bigr]^2\Bigr\},
\label{eq:s2volume_azi_ann}
\end{eqnarray}
will produce an estimate of $\sigma_V^2/\sigma_l^2$ that is appropriate when $V$ is a
cylinder of radius $R$ and height $l$ (equation~\ref{eq:s2volume_azi})
or a cylindrical annulus of inner radius $R_i$, outer radius $R_o$, and height $l$
(equation~\ref{eq:s2volume_azi_ann}).  Since the time-delay
surface (\S~\ref{sec:delay}) has 
rotational symmetry about the $z$ axis, cylindrical
bins are a natural choice for the analysis.

\begin{figure}[htb]
\centerline{\epsfxsize=9cm\epsffile{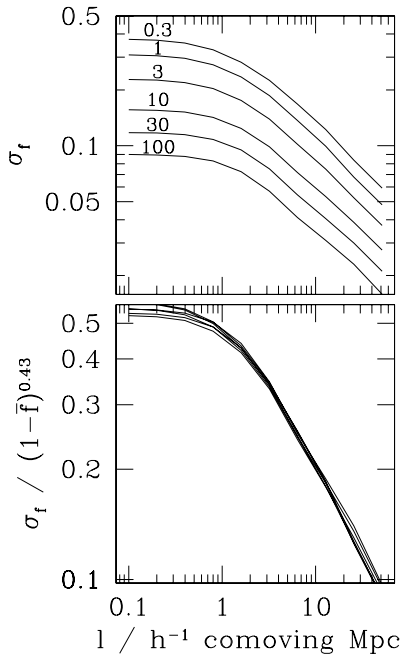}}
\figcaption[f7.eps]{
Top panel: $\sigma_l(l_z)$ (the r.m.s.~transmissivity fluctuation along
a line segment of length $l_z$) recalculated after dividing the observed
Lyman-$\alpha$ optical depths in the QSOs HS1946+7658, 
Q0636+680, Q0956+122, Q0302-003, and Q0014+813 by the number $b$
written next to each curve.  Bottom panel:  same as top, except the 
dependence of the amplitude on the mean transmissivity
$\bar f(b)$ has been crudely removed by dividing
each curve by $(1-\bar f(b))^{0.43}$.  The differences in shape revealed by
this panel are small over the distance range $1\simlt l\simlt 10h^{-1}$ comoving
Mpc that is most important to our analysis.
\label{fig:s2l_vs_b}
}
\end{figure}

A more sophisticated treatment
would take into account the changes to the power-spectrum that
will accompany changes in the ionizing radiation intensity.
That will likely require numerical simulations and is beyond the scope of this
paper.
However, since changes in the radiation field alter the neutral fraction but
not the temperature of intergalactic gas (see \S~\ref{sec:preliminaries}),
they will have a stronger effect on the amplitude of the powerspectrum
than on its shape.  Amplitude-independent results (e.g., our estimate of
$\sigma_V^2/\sigma_l^2$) may not be disastrously affected.
Figure~\ref{fig:s2l_vs_b} presents some evidence in favor of this assertion.
I converted published Lyman-$\alpha$ Voigt profile lists for the QSOs HS1946+7658 (Kirkman \& Tytler 1997,
$z=2.994$),
Q0636+680, Q0956+122, Q0302-003, and Q0014+813 (Hu et al. 1995, $z=3.180$, 3.288,
3.294, 3.366) into 
a list of each QSO's Lyman-$\alpha$ optical depth as a function of redshift,
divided all the optical depths by a constant $b$ to mimic a change
in the ionizing radiation intensity, then calculated the dependence
of $\sigma_l$ on $l$ for various values of $b$.  The top panel of 
figure~\ref{fig:s2l_vs_b} shows the result.  The amplitude of $\sigma_l$
changes significantly with $b$, but its shape, which is sensitive to
the shape of the powerspectrum, does not.  This can be seen more
clearly in the bottom panel, where the change in amplitude has
been crudely removed by scaling each curve according to the relationship
\begin{equation}
\sigma_l \propto [1-\bar f(b)]^{0.43}
\label{eq:s2l_vs_b}
\end{equation}
where $\bar f(b)$ is the mean transmissivity 
after dividing the actual optical depths by $b$.
Once this scaling is removed, the $\sigma_l$ curves have nearly the
same shape over the range $1\simlt l\simlt 10h^{-1}$ comoving Mpc
that is most important in our calculations.

\subsection{Commentary}
\label{sec:commentary}

It is worth considering the relative sizes of the terms in
equation~\ref{eq:s2fbar} before we move on.
Suppose for concreteness that we are interested in measuring changes
in the QSO's luminosity over $10$Myr time-scales, so we have
estimated the mean transmissivity in bins of depth $l_z\sim 1.5$ proper Mpc,
and suppose further that the radius $R$ of our bins was chosen
to fill the $40'$ field-of-view of a mosaicked CCD camera,
$R\sim 10$ proper Mpc at $z=3$ for
$\Omega_M=0.3$, $\Omega_\Lambda=0.7$, $h=0.65$.
In comoving units the bin has $R=26h^{-1}$ Mpc and $l_z=3.9h^{-1}$ Mpc,
which implies $\sigma_l\simeq 0.17$ (figure~\ref{fig:s2l_vs_l}, for $\bar f=0.67$),
$\sigma^2_l/\sigma^2_V\simeq 20$, and consequently
$\sigma_V\simeq 0.04$.  Inserting these numbers into
equation~\ref{eq:s2fbar} leads to the two primary conclusions
of this section:

(A) The spectra of the background objects do not need to be very good.
The second and third terms in equation~\ref{eq:s2fbar} contribute
equally to the total uncertainty when
the signal-to-noise ratio in a spectral segment of length
$l_z\sim 3.9h^{-1}$ comoving Mpc ($\sim 7$\AA) is ${\cal S}=\bar f/\sigma_l \sim 4$.
Obtaining better spectra cannot change the total uncertainty by much.
Random variations in the intergalactic matter density
make the mean HI absorption $\bar f_1$ along a single sightline segment 
a poor indicator of the ionizing background $J_\nu$ and
consequently there is no need to measure $\bar f_1$ with
exquisite precision.  

(B) The ultimate limit on our uncertainty is set by $\sigma_V$ 
and it is a limit that we will reach very quickly.
For the example considered here, the first and second terms
in equation~\ref{eq:s2fbar} have the same size
for $N=20$.  Increasing $N$ further can reduce the second and third
terms arbitrarily but not the total uncertainty.
In practice one will want to obtain samples many times
larger than $N=20$ to help address the possibility
that the QSO's radiation is beamed.
As a result $\sigma_V$ will probably
be the only significant contributor to $\sigma_f$
in realistic cases.
The bottom panel of figure~\ref{fig:fluxfigs} shows
that $\sigma_V$ is large enough to prevent
us from detecting all but the coarsest changes in
a QSO's luminosity.  
We can do nothing about this.
Only a small region of the universe is bathed in the light that
the QSO emitted at one period in its history;
the mean density in this region will stray from the global mean
to the extent that inflation requires; 
the intensity of the QSO's ionizing radiation cannot be measured
if it influences
the region's mean HI absorption by less.

Although the numbers quoted above depend on our arbitrary choice
for the bin size,
neither $\sigma_l$ nor $\sigma_V$ is a strong enough
function of $R$ or $l$ for the qualitative conclusions to change
significantly as the bin size varies across its useful range.
The change of variance with mean transmissivity (equation~\ref{eq:s2l_vs_b})
might seem more important.  If the QSO were bright enough
to drive the mean transmissivity in a bin to $\bar f\sim 0.9$, for example, rather than
the value $\bar f\sim 0.67$ assumed in the preceding three paragraphs,
$\sigma_l$ would decrease to $0.10$, $\sigma_V$ would decrease to $0.02$,
and one would find that spectra of slightly higher S:N were desirable.
In practice, however, these high transmissivities are unlikely to be reached
anywhere except very near the QSO, and here the decrease in $\sigma_l$ and
$\sigma_V$ is largely offset by the small bin sizes that are required at small radii.
This can be seen in the worked example of \S~\ref{sec:synthesis}, below. 

One qualification should be added to my claim (A) that it is a waste of
time to obtain high S:N spectra.  That is necessarily true
only if one is committed to using the mean transmissivity $\bar f$ as
a probe of the radiation density $J_\nu$.  At sufficiently high signal-to-noise ratios,
however, other options are available.  Consider an arbitrary function $g$
of the line-of-sight HI density that has variance $\sigma_g^2$ due
to random fluctuations when
the brightness $b\equiv J_\nu/J_{\rm bg}$ of the ionizing radiation field
is kept constant.  The random fluctuations in $g$ will prevent us from detecting
relative changes in $b$ of order $\Delta{\rm ln}b\sim |d\,{\rm ln}b/d\,{\rm ln}g|\sigma_g/g$.
If $g$ is the mean transmissivity $\bar f$ on a $3.9h^{-1}$ comoving Mpc 
line segment, which is the only possibility we have treated so far, then 
$d\,{\rm ln}b/d\,{\rm ln}g\sim 5$ (near $g=0.67$) and
$\sigma_g/g\sim 0.26$, so the minimum detectable fluctuation in $b$
will have $\Delta{\rm ln}b\sim 1.3$.  At high enough S:N it would be
possible abandon $\bar f$ and adopt (say) the total HI column-density 
${\cal N}$ along the
same line segment as the probe of $b$.  In this case, with $g={\cal N}$,
we have $d\,{\rm ln}b/d\,{\rm ln}g\simeq -1$ and $\sigma_g/g \sim 6$, where
$\sigma_g/g$ is the value that obtains among the 5 QSOs with $z\sim 3$
discussed at the end of \S~\ref{sec:arithmetic}, so the
minimum detectable fluctuation in $b$ has $\Delta{\rm ln}b\sim 6$.
${\cal N}$ is evidently far inferior to $\bar f$ as an estimator of $b$.
A better choice, letting $g$ equal $n$, the number of detected Lyman-$\alpha$ forest lines
on the segment, was advocated by Bajtlik, Duncan, \& Ostriker (1988).
If the column density distribution has the form 
$P(N_{\rm HI})\propto N_{\rm HI}^{-\alpha}$, then 
$d\,{\rm ln}b/d\,{\rm ln}g\simeq (1-\alpha)^{-1}$.  Among the same 5 QSOs,
$\sigma_n/n\simeq 0.5$ for line segments of length $3.9h^{-1}$ comoving Mpc,
so the minimum detectable fluctuation in $b$ has $\Delta{\rm ln}b\sim 1$ 
for the observed slope $\alpha\sim 1.5$.  Surprisingly, $n$ is only
marginally better than $\bar f$ as an estimator of $J_\nu$.  The
mean transmissivity $\bar f$ along a segment of a
low S:N galaxy spectrum can provide almost as strong a constraint on
the intensity of the QSO's radiation as a high S:N spectrum
analyzed with the standard approach of Bajtlik, Duncan, \& Ostriker (1988).
It would be interesting to extend this analysis from $\sigma_l$
to the more relevant quantity $\sigma_V$.  Since the ratio $\sigma_l/\sigma_V$
depends on the powerspectrum, and different estimators $g$ will have
different powerspectra, it is not necessarily true that the best estimator
for $\sigma_l$ will be best for $\sigma_V$.  This method
would be made far more powerful if one could finding an estimator that
significantly reduces $\sigma_V$.

\section{FEASIBILITY}
\label{sec:feasibility}
The previous section glibly claimed that observational uncertainties
can be made smaller than cosmic variance.
This section considers
the claim in more detail.  
The optimal depth for the spatial bins 
is $\sim 7$\AA, corresponding roughly to 10 Myr time resolution
(see \S~\ref{sec:binsize}), and for the fields-of-view considered
here the cosmic variance will consequently be $\sigma_V\sim 0.04$
(see \S\S~\ref{sec:uncertainties} and~\ref{sec:synthesis}).
This is the level to which we must reduce the
observational uncertainty $\sigma_{\rm noise}$.

\subsection{Signal}
\begin{figure}[htb]
\centerline{\epsfxsize=9cm\epsffile{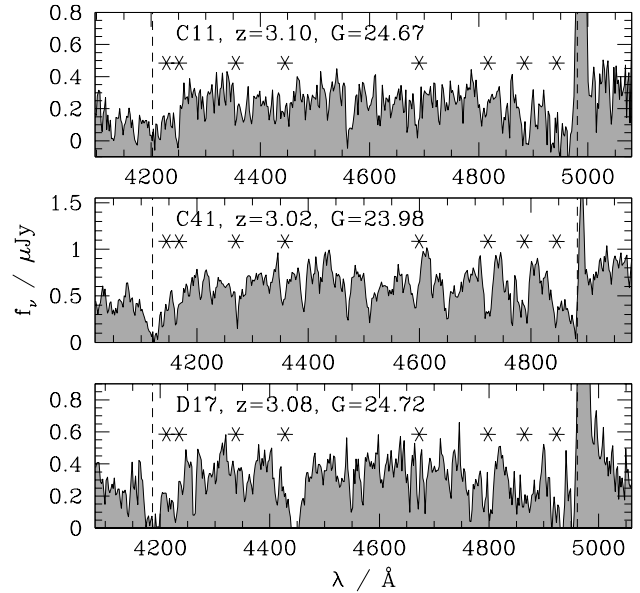}}
\figcaption[f8.eps]{
Examples of the Lyman-$\alpha$ forest in the spectra of galaxies
at redshift $z\sim 3$.  The data were taken with LRIS-B (Steidel et al. 2004)
and were generously provided by A. E. Shapley.  Redshifts and AB magnitudes
are indicated; the names (``C11'' etc.) refer to the SSA22a catalog
of Steidel et al. (2003).
Asterisks mark the locations of the interstellar
absorption lines discussed in figure~\ref{fig:shapley}, below.
Dashed lines indicated the wavelengths of Lyman-$\alpha$ and Lyman-$\beta$
at the redshift of the galaxies' interstellar absorption lines.
\label{fig:threedeepgals}
}
\end{figure}

Reducing the random errors to the desired level $\sigma_{\rm noise}\sim 0.04$ 
is not
much of a challenge.  Figure~\ref{fig:threedeepgals} shows spectra with $\sim 10$\AA\ resolution
of three galaxies in the field SSA22a (Steidel et al. 2003) 
that were observed for $\sim 29000$s with the blue LRIS spectrograph (Steidel et al. 2004)
on the Keck I telescope.  The redshifts and AB-magnitudes of the galaxies
($z\sim 3.1$ and $G\sim 24.5$, respectively) are shown on the plot.
The spectra are preliminary reductions 
that were selected more-or-less at random from the sample
of Shapley et al. (2004, in preparation).
These spectra have signal-to-noise ratios 
in the Lyman-$\alpha$ forest, for
7\AA\ bins ($\sim 1.5$ proper Mpc), of around~5--6.
Averaging together $\sim 10$ spectra of similar
quality would reduce the random errors to the
desired level.  Roughly forty spectra would be
required if the exposure time were 2 hours instead of 8.

\subsection{Continuum fitting}
Systematic errors in the continuum fitting are a source of greater
concern.  We are interested not in a spectrum's flux itself
but rather in its implied Lyman-$\alpha$ transmissivity,
which is the ratio between the observed flux and the continuum flux,
i.e., the flux that
would have been observed in the absence of Lyman-$\alpha$ absorption.
The uncertainty in a bin's mean transmissivity therefore
has an additional term that arises from errors in the estimated continuum.

The size of these errors depends on the method that is used
to estimate the continuum level.
Traditional methods are poorly suited
to the present case; they exploit 
the occasional presence of spectral regions 
with little Lyman-$\alpha$ absorption, and
these regions are rare and difficult to recognize in noisy,
low-resolution galaxy spectra.  
Several authors (e.g., Hui et al. 2001) have presented alternatives
that are more useful to us.  Particularly simple is the method of
Croft (2004), in which the 
continuum level $c(\lambda)$ is estimated by simply smoothing each object's
spectrum and scaling appropriately:
$c(\lambda)\simeq f_{50}(\lambda)/\bar f(z)$, where $f_{50}(\lambda)$ is the object's 
Lyman-$\alpha$ forest spectrum smoothed by a Gaussian with
$\sigma=50$\AA\ and $\bar f(z)$ is the published mean transmissivity 
at redshift $z=\lambda/\lambda_{{\rm Ly}\alpha} - 1$,
which has been calculated by other authors from more-sophisticated continuum fits
to the observed spectra of numerous bright QSOs.

\begin{figure}[htb]
\centerline{\epsfxsize=9cm\epsffile{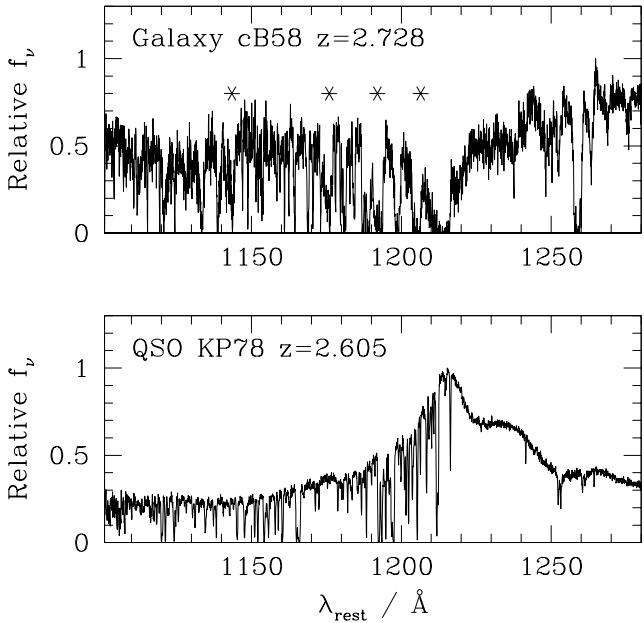}}
\figcaption[f9.eps]{
Comparison of galaxy and QSO continua in the Lyman-$\alpha$
forest region.  The top panel shows the gravitationally lensed
galaxy MS1512-cB58 (Pettini et al. 2002); the bottom
panel shows the QSO Q1623 KP78 (Adelberger et al. 2004, in preparation).
Both spectra were taken with the same instrument, though the
signal-to-noise ratio for the bright QSO is significantly higher.
Asterisks in the top panel mark the locations of the interstellar
absorption lines discussed in figure~\ref{fig:shapley}, below.
\label{fig:cb58kp78}
}
\end{figure}

\begin{figure}[htb]
\centerline{\epsfxsize=9cm\epsffile{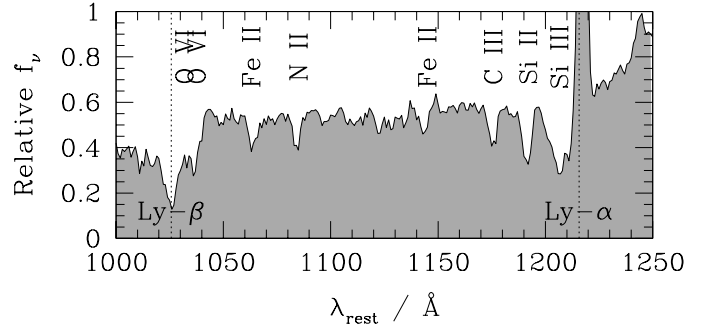}}
\figcaption[f10.eps]{Mean absorption vs. wavelength for galaxies
at redshift $z\sim 3$.  The shaded curve shows the mean
spectrum of 811 galaxies at $z\sim 3$; the data are taken from
Shapley et al. (2003).  Lyman-$\alpha$ forest fluctuations
cancel out in this average, producing a smooth shelf 
with mean transmissivity $\bar f\sim 0.67$ between
Lyman-$\alpha$ and Lyman-$\beta$.  Absorption lines in this shelf
are produced by material intrinsic to the galaxies.  This interstellar absorption
can be safely ignored when its rest-frame equivalent width
is less than $\sim 0.3$\AA (see text).  Stronger lines
have been marked with vertical labels naming the ion responsible
for the absorption.  Data at these wavelengths should be excluded
when calculating the mean transmissivity in different spatial bins.
Note that the resolution of this spectrum ($\sim 12$\AA) is somewhat
lower than the 7\AA\ resolution advocated in the text.
\label{fig:shapley}
}
\end{figure}

Experimentation on the 7 QSOs with $z\sim 3$ in the sample of
Adelberger et al. (2003) shows that transmissivities estimated
with Croft's (2004) approach and the traditional approach are very 
similar.  Excluding the parts of the Lyman-$\alpha$ forest that fall
on the QSOs' Lyman-$\alpha$ and Lyman-$\beta$ emission lines,
the correlation coefficient 
between the transmissivities estimated with the two approaches is
$r\sim 0.96$--$0.98$. This implies that the r.m.s.~difference
between them, $(1-r^2)^{1/2}\sigma_l$, is $\sim 0.03$--$0.05$ 
for 7\AA\ bins with $\sigma_l\simeq 0.17$ (see Figure~\ref{fig:s2l_vs_l}).
Since galaxies' and QSOs' continua between Lyman-$\alpha$ and Lyman-$\beta$ 
are similarly featureless (see, e.g., figures~\ref{fig:threedeepgals},~\ref{fig:cb58kp78},
and~\ref{fig:shapley}), errors in continuum fitting should introduce
a similar uncertainty in galaxies' Lyman-$\alpha$ forest transmissivities.
Even if these errors were correlated from
one galaxy to the next, and did not tend to cancel in the averaged
transmissivity in each spatial bin, their size would be no larger than
the cosmic variance.  In practice continuum errors from the
Croft (2004) approach should cancel significantly,
however.  They arise, to a large extent, because the Lyman-$\alpha$
forest is not completely uniform on $\sim 100$\AA\ scales and 
any deviations from uniformity are incorrectly interpreted as features
of the continuum.  The resulting r.m.s.~continuum error 
averaged over a spatial bin can be calculated with the approach
of~\S~\ref{sec:uncertainties}:
\begin{equation}
\sigma_{\rm cont} \simeq \biggl[\frac{N-1}{N}{\sigma'}_V^2 + \frac{1}{N}{\sigma'}_l^2\biggr]^{1/2}
\end{equation}
where ${\sigma'}_V^2$ and ${\sigma'}_l^2$ are given by 
equations~\ref{eq:s2volume} and~\ref{eq:s2line} with
$\sin(k_zl_z/2)/(k_zl_z/2)$ replaced by $\exp[-k_z^2\sigma_z^2/2]$,
where $\sigma_z$ is the comoving distance corresponding to
the $50$\AA\ smoothing length.  Since $\sigma_z\gg l_z$, 
${\sigma'}_V^2$ will be smaller than $\sigma_V^2$,
${\sigma'}_l^2$ will be smaller than $\sigma_l^2$, 
and the uncertainty from this source of continuum errors will 
never dominate the total uncertainty.  This is true even in the 
low signal-to-noise
limit, since in this case the uncertainty in the smoothed continuum will
be dwarfed by the uncertainty in a single $\sim 7$\AA\ bin.


\subsection{Interstellar absorption lines}
Although galaxies' continua appear to be mostly featureless
between Lyman-$\alpha$ and Lyman-$\beta$, there are some
important exceptions:  at a handful of wavelengths
the galaxies' interstellar absorption lines are too strong to
be ignored.  These wavelengths are apparent in figure~\ref{fig:shapley},
which shows average observed absorption in
region between Lyman-$\alpha$ and Lyman-$\beta$
in a sample of 811 Lyman-break galaxies (Shapley et al. 2003).
Since the r.m.s.~variation in Lyman-$\alpha$ forest transmissivity
in 7\AA\ bins is $\sigma_l\simeq 0.17$ (Figure~\ref{fig:s2l_vs_l}),
interstellar absorption lines will have a non-negligible effect
on the estimated transmissivity if their rest-frame equivalent width
exceeds $7$\AA$\times\sigma_l/(1+z)\simeq 0.3$\AA.  In the mean spectrum
of Shapley et al. (2003) there are 7 such lines between
Lyman-$\alpha$ and Lyman-$\beta$.  The existence of these
interstellar absorption lines will not have a disastrous effect
on the analysis, since each background galaxy will lie at a slightly different
redshift, but one might as well eliminate their effect completely by
masking the relevant portions of the spectrum.  This will reduce
the effective number of background galaxies by too modest an amount
to significantly alter the significance of any conclusions.

\section{SYNTHESIS}
\label{sec:synthesis}

\begin{figure}[htb]
\centerline{\epsfxsize=9cm\epsffile{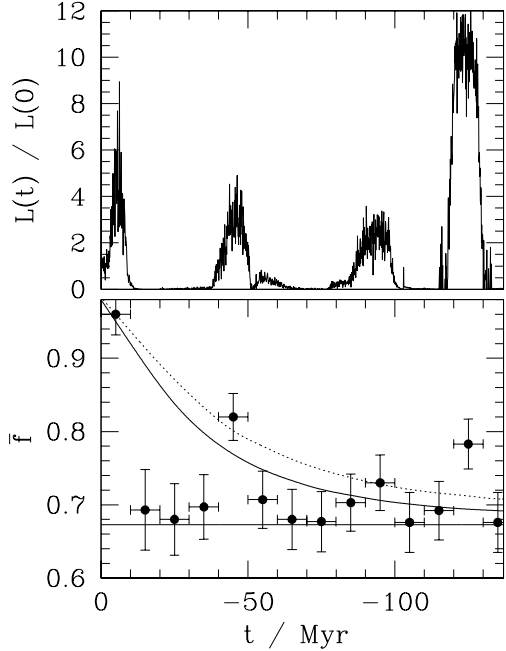}}
\figcaption[f11.eps]{
Top panel:  the QSO's ionizing luminosity as a function of time.
This curve, generated for the example of section~\ref{sec:synthesis},
is not intended to be a realistic model of a QSO's output.  It features
were chosen to illustrate a number of points made in the text.
Bottom panel:  the recovered mean transmissivity in bins whose edges
align with contours of the time-delay surface.  Points mark the
mean transmissivity calculated from figure~\ref{fig:qpedelay_rad};
error bars show the $1\sigma$ range that would be observed if
figure~\ref{fig:qpedelay_rad} included cosmic variance, the dominant
noise source.  The solid curve shows the mean transmissivity than would have
been observed if the QSO's luminosity were constant, $L(t)=L(0)$.
The dotted curve shows the mean transmissivity in the primed ``control'' bins
(figure~\ref{fig:qpedelay_rad}); it differs from the solid line due to the
increase in the QSO's luminosity from $t=0$ to $t=-10$, though this
effect would presumably be removed by a maximum likelihood fit of the data
to a surface.
Decreases of the luminosity to 0 and subsequent increases back to $L(0)$ can
be detected with moderate significance if they happened within the interval
$-50\simlt t\simlt 0$ Myr.  Only extremely large increases in the QSO's luminosity
can be detected at earlier times.
\label{fig:recovered_flux}
}
\end{figure}

We can now estimate how easily we will be able to detect ionization gradients produced
by changes in a QSO's luminosity.  This section works through the details for
a single case, an isotropically radiating QSO of ($t=0$) magnitude $m_{912}=18$
whose ionizing luminosity varied
with time according to the curve $L(t)$ shown in the top panel of figure~\ref{fig:recovered_flux}.
An $\Omega_M=0.3$, $\Omega_\Lambda=0.7$, $h=0.65$ cosmology with the uniform
ionizing background of equation~\ref{eq:jbg}
will be assumed throughout this section.

\begin{figure}[htb]
\centerline{\epsfxsize=9cm\epsffile{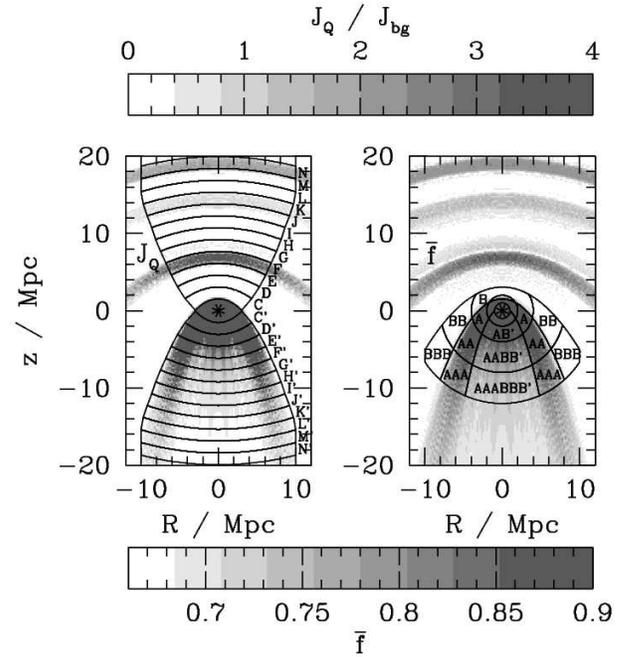}}
\figcaption[f12.eps]{
Left panel:  variations in the intensity of ionizing radiation in the observed
frame for a QSO at redshift $z=3$ with $m_{912}=18$ that has the history of outbursts
shown in figure~\ref{fig:recovered_flux}.  Also shown are sample bins used in the analysis
at larger lookback times.
Right panel:  variations in the expected mean transmissivity implied by the radiation
intensity of the left panel.  The mean transmissivity would hew to these values only
if results from numerous identical
QSOs were averaged.  A more realistic single realization would 
show large fluctuations in the transmissivity due to random variations in the density
of intergalactic matter.  The labeled regions surrounding the QSO
are sample bins for the analysis at smaller lookback times.
\label{fig:qpedelay_rad}
}
\end{figure}

According to equation~\ref{eq:Jr}, the radiation intensity as a function of position 
in the observed frame
(i.e., the radiation intensity that was present when the photons from the background
sources passed through each point) is
\begin{equation}
J(R,z) = \Bigl[1+\frac{L(t_I(R,z))}{L(0)}\Bigl(\frac{r_{\rm eq}}{r}\Bigr)^2\Bigr] J_{\rm bg}
\end{equation}
where $t_I(R,z)$ and $r_{\rm eq}$ are given by equations~\ref{eq:req}
and~\ref{eq:timedelay}.  The left panel of figure~\ref{fig:qpedelay_rad} shows
this function for $r_{\rm eq} = 8.27$ proper Mpc, which is appropriate
to the QSO described in the preceding paragraph.
Since the ionization and recombination times (\S~\ref{sec:preliminaries})
are short compared to the time for significant changes in $L(t)$,
the neutral fraction will be inversely proportional to $J(R,z)$, and the mean corresponding
transmissivity $f(R,z)$ can be derived from the curve shown in the top panel
of figure~\ref{fig:fluxfigs}.  The result is shown in the right panel of 
figure~\ref{fig:qpedelay_rad}.  The panel shows the mean transmissivity
that would be observed if one averaged results from many identical QSOs;
the actual transmissivity surrounding a single QSO would
have significant variations about this mean, due primarily to random
fluctuations in the density of intergalactic matter.   

Changes in the QSO's ionizing luminosity could be detected in various ways, but
for now I will assume that one is
aiming to detect the changes
by looking for differences in the mean transmissivity $\bar f_{\rm bin}$ among bins whose edges
trace contours of the time delay surface $t_I(R,z)$.  One set of such bins is shown
in the left panel of figure~\ref{fig:qpedelay_rad}.  Also indicated are symmetrically distributed
``control'' bins.  These bins have shapes identical to the others, but are located on
the opposite side of the QSO, in a region that is illuminated by the
QSO's radiation at short time delays $0>t>-10$ Myr.  The mean transmissivity in these bins
provides an indication of how the mean transmissivity for $z>0$ would vary with position
if the QSO's luminosity were always equal to its observed ($t=0$) value(\S~\ref{sec:delay}).
Note that the requirement
$t>-10$ Myr for the control bins limits the size of their partner bins at smaller radii.

Our ability to detect changes in the QSO's luminosity will depend on the uncertainty
in the binned transmissivity $\bar f_{\rm bin}$.
For a reasonable number of background
sources ($\simgt$ few dozen) this uncertainty will be nearly
equal to the 
cosmic variance $\sigma_V$ (\S~\ref{sec:uncertainties}).  As discussed
in \S~\ref{sec:uncertainties}, the size of $\sigma_V$ can be
roughly estimated by (1) approximating each bin as a cylinder of  radius $R_{\rm bin}$ and depth
$l_{\rm bin}$, (2) calculating r.m.s.~transmissivity fluctuation along a skewer of length
$l_{\rm bin}$ by interpolating from figure~\ref{fig:s2l_vs_l} and scaling according to 
the local expected transmissivity $f(R,z)$ with equation~\ref{eq:s2l_vs_b},
and (3) multiplying by $\beta^{1/2}$, where 
$\beta(R_{\rm bin},l_{\rm bin})\equiv \sigma^2_V/\sigma^2_l$ is
a constant that can be determined for each bin by numerically integrating 
equations~\ref{eq:s2line}, \ref{eq:s2volume_azi}, and~\ref{eq:pkmcd}.
The bottom panel of figure~\ref{fig:recovered_flux} shows the expected mean transmissivity in
each bin along with the estimated $1\sigma$ uncertainty calculated in this way.

\begin{figure}[htb]
\centerline{\epsfxsize=9cm\epsffile{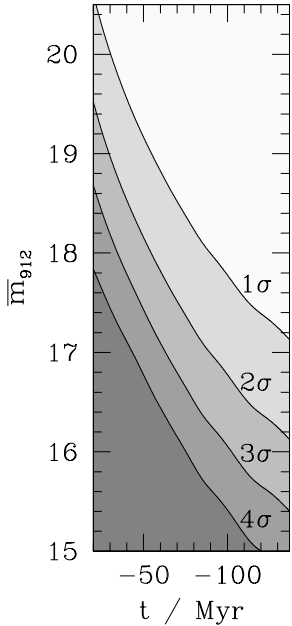}}
\figcaption[f13.eps]{
The $10$Myr running-average luminosity required at various lookback times for
the flux in the corresponding $10$Myr bin to be $N\sigma$
above the background level.   Luminosity changes among
even the brightest QSOs will be difficult to detect 
at moderate significance for $t\simlt -100$ Myr.
\label{fig:mlim_vs_t}
}
\end{figure}

The main conclusions from figure~\ref{fig:recovered_flux} are (1) that
only large changes in the QSO's luminosity will leave an imprint on the IGM
that is easily detectable with this approach, and (2) that the minimum size of
a detectable luminosity difference increases rapidly towards earlier times.
These conclusions are easier to apprehend in figure~\ref{fig:mlim_vs_t},
which shows the 10-Myr moving-average magnitude $\bar m_{912}$ a QSO would need 
to have had at various
lookback times to for its sudden death (or revival) to leave a detectable
fossil record in the spectra of background galaxies.  Measuring changes in a QSO's
luminosity will be difficult for $t\simlt -50$ Myr and nearly impossible
for $t\simlt -100$ Myr.  At smaller time delays the prognosis is good.

\begin{figure}[htb]
\centerline{\epsfxsize=9cm\epsffile{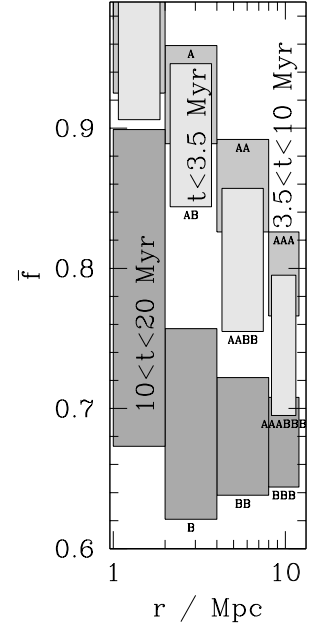}}
\figcaption[f14.eps]{
The mean transmissivity and its uncertainty in the bins shown in
the right panel of figure~\ref{fig:qpedelay_rad}.  Bins with the
lightest shading are bathed in light emitted by the QSO at
$-3.5{\rm Myr}<t<0$; bins with medium shading by light emitted
at $-10<t<-3.5$Myr; bins with darkest shading by $-20<t<-10$Myr.
The height of the bins corresponds to the mean transmissivity $\pm 1\sigma$.
Bin names in this figure correspond to those in
figure~\ref{fig:qpedelay_rad}.  Bins with $1<r<2$Mpc are unlabeled
in figure~\ref{fig:qpedelay_rad} for clarity.  At fixed time delay,
$\bar f$ declines with radius due to the $1/r^2$ decline in flux from
the QSO.  Comparing $\bar f$ radius by radius clearly reveals the decline
in $L(t)$ (figure~\ref{fig:recovered_flux})
from $-10{\rm Myr}<t<0$ to $-20<t<-10$Myr.  The increase from
$-3.5{\rm Myr}<t<0$ to $-10<t<-3.5$Myr would not be easy to detect;
a decrease during the same times would have been detected with high
significance.
\label{fig:recovered_flux.smallr}
}
\end{figure}

As $t\to 0$, the shape of the time-delay contours becomes increasingly
incompatible with the goal of having symmetrical control bins.
For $-20\simlt t\simlt 0$ the best approach may be to abandon
the binning altogether and simply find the maximum likelihood fit of the data
to an appropriate surface.  To give some indication of the uncertainty
in the radiative history that would result, however, I will 
continue a binned analysis with
the bins shown in the right panel of figure~\ref{fig:qpedelay_rad}.
The outermost bins ($B$--$BBB$) enclose regions with time delays
$-20<t<-10$ Myr, the middle bins ($A$--$AAA$) enclose 
$-10<t<-3.5$ Myr, and the innermost ($AB'$ etc.) enclose $-3.5<t<0$ Myr.
A complication for these small time delays is that a region
with fixed $t$ has a wide range of distances $r$ to the QSO.
Variations in $\bar f$ due to changes in $1/r^2$ could obscure 
the variations of interest from changes in the QSO luminosity $L(t)$.
My approach in this simplified analysis is to divide each time-delay region
into a number of bins with similar values of $r$ (e.g., bins
$B$, $BB$, and $BBB$ for $-20<t<-10$ Myr).  Figure~\ref{fig:recovered_flux.smallr}
shows the mean transmissivity within each of these bins.  The uncertainty
in the bins was calculated by approximating them
as cylinders ($B$, $AB'$, $AABB'$, $AAABBB'$) or cylindrical annuli
($A$--$AAA$, $B$--$BBB$) with the same volume and roughly the 
same shape; the expected variance does not depend sensitively
on the details of this approximation.

With these bins, 
the decrease
in $L(t)$ from $-10<t<0$ Myr to $-20<t<-10$ Myr is detected with
some significance at each radius and with high significance
when the results from different radii are combined.  If the decrease
had happened at an earlier time $-10<t\simlt -3.5$ Myr it would have
been detected with higher significance still.  Instead the
luminosity increased from $-3.5{\rm Myr}<t<0$ to $-10<t\simlt -3.5$.
As the plot shows, increases in the luminosity at small look-back times
are much harder to detect than decreases, since the sensitivity
of $\bar f$ to $J_Q$ falls as $\bar f\to 1$ (figure~\ref{fig:fluxfigs}).
Fortunately decreases must be far more likely than increases:
QSOs with $m_{912}=18$ lie on the steep bright-end of the luminosity function.

\section{LIMITS TO THE TIME RESOLUTION}
\label{sec:binsize}
In the previous sections the data were placed into spatial bins
whose depth $l_z=1.5$ Mpc gave us sensitivity to luminosity fluctuations
time-scales of $10$ Myr or greater.  Ideally one would be able to
detect fluctuations 
on any time scale. 
Could we have achieved significantly better time resolution
by placing the data in bins with smaller $l_z$?  The answer is no;
this section explains why.

If cosmic variance were the only problem,
the bin depth $l_z$ could
be made arbitrarily small.
$\sigma_V$ 
(equation~\ref{eq:s2volume_azi}) 
is almost independent
of $l_z$ for $l_z\ll R$, the case of interest, and so a bin
that is infinitesimally thin will have nearly the same
cosmic variance as the adopted bins with $l_z=1.5$ proper 
Mpc.\footnote{The reason is that a cylinder with $l_z\ll R$
in real space will have $l_z\gg R$ in Fourier space,
and since the power $P(k)$ is concentrated near $k\sim 0$
the variance is dominated by contributions from wavenumbers
${\mathbf k}$ near the origin.  There is little power
at the distant ends of the cylinder with large $k_z$, 
and altering the limits of the $k_z$ integral in
equation~\ref{eq:s2volume_azi} (i.e., changing
the thickness $l_z$ of the cylinder in real space) does not
affect the variance by much.
}

Unfortunately the ability to obtain reasonably accurate estimates of
the mean transmissivity
in infinitesimally thin bins
is not the same as the ability to measure changes in the
QSO's luminosity that happened on arbitrarily short time scales.
Our estimate of a gas element's longitudinal separation $z$ from the QSO 
will be inaccurate for two reasons: (1) we will not know the
precise redshift of the QSO, and (2) our $z$ positions
are derived from redshifts and will be distorted from their true values
by peculiar
velocities.  As a result the time delay to the element is uncertain.
When the time delays to different elements are uncertain, we cannot
combine elements with exactly the same delays into one bin;
the various elements that make up a single bin will inevitably
have a range of time delays.  The minimum range of time delays in a bin
is what limits our time resolution.  The remainder of the section 
considers this limit in more detail.

\subsection{Uncertainty in the QSO redshift}
If the QSO's redshift is measured from the CIV emission line, 
the uncertainty in its
systemic recession velocity will be $\sigma_v\simeq 510 {\rm km}\,{\rm s}^{-1}$
(Richards et al. 2002), which corresponds to a positional
uncertainty of $\sigma_z\simeq 1.75 h_{65}^{-1}$ proper Mpc
at $z=3$ for $\Omega_M=0.3$, $\Omega_\Lambda=0.7$.
The uncertainty can be reduced to $0.9 h_{65}^{-1}$ proper Mpc
(i.e., $270 {\rm km}\,{\rm s}^{-1}$) if MgII is used instead (Richards et al. 2002),
and to $0.3 h_{65}^{-1}$ proper Mpc (i.e., $80 {\rm km}\,{\rm s}^{-1}$)
if [OIII] is used (Vrtilek \& Carleton 1985).
Radio observations of molecular emission lines could presumably reduce
$\sigma_z$ even further, but this is unlikely to benefit us much.
The time resolution achievable with $\sigma_z=0.3$ proper Mpc
is $t\sim 2\sigma_z/c\sim 2$ Myr, and other effects prevent us from
obtaining a resolution even this coarse.

\subsection{Thermal motions}
A firm lower limit to the time resolution is set by the thermal
motions of the $20000$ K intergalactic gas.  The intergalactic
hydrogen at a particular
true $z$ position will have an rms range of apparent $z$ position
of $\sigma_z = (kT/m_H)^{1/2}/H \simeq 0.04$ proper Mpc
for $\Omega_M=0.3$, $\Omega_\Lambda=0.7$, $h=0.65$.
We will not be able to measure changes in the QSO's
luminosity that happen much more rapidly than the corresponding time scale
$t\sim 2\sigma z/c\sim 0.3$ Myr.  This is unlikely to be the limiting factor
in the analysis.

\subsection{Streaming towards the QSO}
\begin{figure}[htb]
\centerline{\epsfxsize=9cm\epsffile{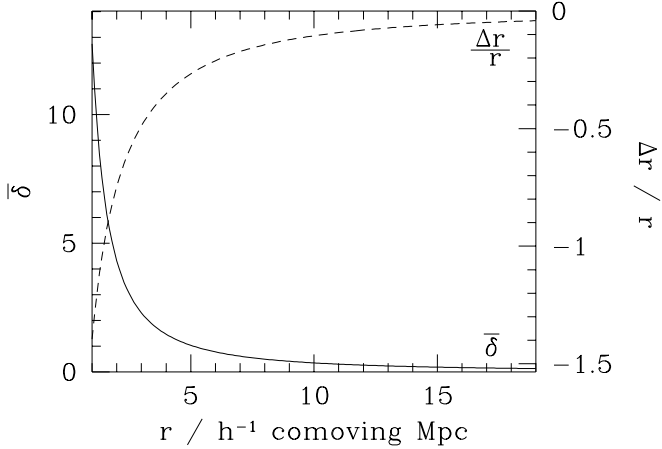}}
\figcaption[f15.eps]{The mean interior overdensity (solid line) and
approximate infall velocity (dashed line) as a function of distance
from a bright QSO at $z=3$.  The infall velocity is expressed
as $\Delta r/r \equiv (r_{\rm apparent}-r_{\rm true})/r_{\rm true}$, the
relative shift in an object's position that the infall velocity
would produce if the velocity were directed exactly towards the observer.
The infall velocity is estimated with a simple version of the 
Zeldovich approximation
that becomes inaccurate for $\bar\delta\gg 1$; see text.
\label{fig:vzeldo}
}
\end{figure}

The effect of larger-scale peculiar velocities is more severe.
First there is the average streaming motion towards the QSO, which can
be crudely estimated as follows.  
If the scale dependence of QSOs' bias $b$ is weak,
then the mean matter overdensity at a distance $r$
from a QSO will be roughly $\delta(r) = \xi_Q(r)/b$,
where $\xi_Q$ is the correlation function of QSOs.
According to Croom et al. (2002), a correlation function
of the form $\xi_Q(r) = (r/r_0)^{-\gamma}$, $r_0\simeq 8.4 h^{-1}$ comoving
Mpc, $\gamma\simeq 1.56$ is appropriate for the brightest QSOs
at any redshift.  The variance of QSO number density in cells
of radius $r_{\rm cell}=8 h^{-1}$ comoving Mpc is therefore
$\sigma_Q^2 = 72(r_0/r_{\rm cell})^\gamma / [(3-\gamma)(4-\gamma)(6-\gamma)2^\gamma]\simeq 1.3$,
(Peebles 1980 eq. 59.3),
which implies a QSO bias at $z=3$ of $b\simeq 4.5$ if $\Omega_M=0.3$, $\Omega_\Lambda=0.7$,
and the rms linear matter-density fluctuation in $r=8 h^{-1}$ Mpc spheres
at $z=0$  is $\sigma_8=0.9$.
Integrating over the correlation function shows
that the mean matter overdensity within a comoving radius $r_c$, $\bar\delta(r_c)$, 
is roughly 
\begin{equation}
\bar\delta(r)\simeq \frac{3r_c^{-\gamma}}{(3-\gamma)br_0^{-\gamma}}.
\end{equation}
Now
the proper radius $r_p$ of a spherically symmetric
region with mean interior overdensity $\bar\delta$ evolves according
to
\begin{equation}
\frac{\dot r_p}{r_p} = H - \frac{1}{3(1+\bar\delta)}\frac{d\bar\delta}{dt}
\label{eq:apsphcol}
\end{equation}
where $H(t)$ is the Hubble parameter.  This follows from the fact
that concentric shells of matter do not cross until just before
final collapse.
Adopting the spherical Zeldovich approximation for simplicity, the linear overdensity
density $\bar\delta_L$ will be related to the true overdensity through
$1+\bar\delta \sim (1-\bar\delta_L/3)^{-3}$, which reduces
equation~\ref{eq:apsphcol}
to
\begin{equation}
\frac{\dot r_p}{r_p} = H\Bigl[1+f-f\,\bigl(1+\bar\delta(r_p)\bigr)^{1/3}\Bigr],
\label{eq:apsphzel}
\end{equation}
and shows that
\begin{equation}
\Delta z = f\frac{z}{(R^2+z^2)^{1/2}}\Bigl[1-\bigl(1+\bar\delta(r_p)\bigr)^{1/3}\Bigr],
\label{eq:dzsphzel}
\end{equation}
is the rough proper distance between the a volume element's true position $(R,z)$ and
the position we that we erroneously 
infer from assuming that it is at rest with respect to 
the Hubble flow.  Here $f\equiv d {\rm ln} D / d {\rm ln} a \simeq \Omega_M^{0.6}(z)$ where $D$ is the linear-growth factor and $a$ is the scale factor of the
universe.  Figure~\ref{fig:vzeldo} shows $\bar\delta$ and
$\Delta z$ as a function of distance.

By itself the net streaming
towards the QSO is not a major problem,
at least for $r\simgt 5h^{-1}$ comoving Mpc.
Its primary effect is to make absorbing gas appear
to be closer to the QSO than it actually is.  Although this produces
a slight systematic error in the lookback time assigned to each volume element,
the error can be corrected to a large degree with
simple formulae such as equation~\ref{eq:dzsphzel}, and in any
case slight inaccuracies in the times assigned to the $x$ axis
of figure~\ref{fig:recovered_flux} would not diminish its scientific
value by much.

\subsection{Random peculiar velocities}
\label{sec:randompecvel}
\begin{figure}[htb]
\centerline{\epsfxsize=9cm\epsffile{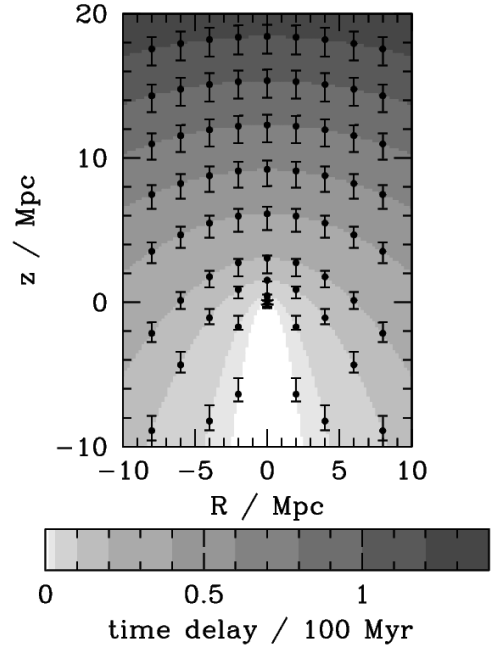}}
\figcaption[f16.eps]{
The peculiar velocity field.  Peculiar velocities
add random offsets to the apparent $z$ positions of intergalactic gas
and systematically shift them towards the QSO.  Black points mark
the true spatial positions of selected intergalactic volume elements;
error bars show the expected $1\sigma$ range of each volume element's
apparent position.  The mean offsets were calculated with 
equation~\ref{eq:dzsphzel} and the sizes of the ranges with
equation~\ref{eq:sig2vz_linear}.  Contours in the background
show the time delay to each position, starting with $1$, $3$, $10$, $20$ Myr
and increasing in steps of 20 Myr thereafter.  Because of peculiar velocities,
gas with a range of time delays is mixed together in any spatial bin
that is defined in the observed (redshift-space) frame.  This limits
the time resolution of the method. 
For time delays
greater than $\sim 20$ Myr, the
time resolution is limited to $\simgt 10$ Myr.  At very small time
delays the resolution is far better. Peculiar velocities are unlikely
to make material with a delay of $\sim 1$ Myr appear to have a delay of 3 Myr,
for example, so we should be able to use the method to detect QSO lifetimes
of order 1 Myr even though we will not be able to distinguish a life time
of 30 Myr from one of 31 Myr.
\label{fig:sigmaz}
}
\end{figure}

More troubling are random deviations around the net streaming
motion.  As a result of them, the gas that is illuminated by
the QSO's luminosity at time $t$ will lie on a complicated
surface that wanders randomly around the parabolic time-delay
surface shown in figure~\ref{fig:qpedelay}.  Since particles maintain
their linear velocities long after the density field itself
has left the linear regime, and since most of intergalactic space should
be occupied by low density (i.e., uncollapsed) gas that is not
far from the linear regime,  linear perturbation
theory should provide a rough estimate of the typical size
of these excursions.  Let $\sigma_{v_z}({\mathbf r})$ be the
rms difference in the $z$ comoving peculiar velocities of two points separated
by the vector ${\mathbf r}$.
In the linear regime the Fourier transform $\tilde{\mathbf v}({\mathbf k})$
of the comoving peculiar-velocity field ${\mathbf v}({\mathbf r})$ is related
to the Fourier transform $\tilde\delta({\mathbf k})$ of the comoving density
field $\delta({\mathbf r})$ through 
$\tilde{\mathbf v}({\mathbf k}) = -iHf\tilde\delta({\mathbf k}){\mathbf k}/k^2$,
(e.g., Peebles 1980 equation 27.22)  
where $H$ is the Hubble parameter and $f\simeq\Omega_M^{0.6}(z)$.
Convolving
$v_z({\mathbf r})$ by the sum of a positive delta function at ${\mathbf r'}/2$
and a negative delta function at $-{\mathbf r'}/2$ produces a new random
field whose value at each point is equal to the $z$-velocity difference
between the points
${\mathbf r'}/2$ and $-{\mathbf r'}/2$.  The variance of this field
is equal to $\sigma_{v_z}^2({\mathbf r})$, which can therefore
be written, according to equation~\ref{eq:s2general}, as
\begin{equation}
\sigma_{v_z}^2({\mathbf r}) = \frac{H^2f^2}{2\pi^3} \int d^3k\,\frac{k_z^2}{k^4} P_L({\mathbf k}) \sin^2\Bigl(\frac{{\mathbf k}\cdot{\mathbf r}}{2}\Bigr).
\end{equation}
Expressing $\sigma_{v_z}^2({\mathbf r})$ in terms of one-dimensional
integrals by
converting to spherical coordinates and integrating over solid angle,
one finds
\begin{equation}
\sigma_{v_z}^2({\mathbf r}) = 2[\psi_{\|}(0)-\psi_{\|}(r)\cos^2(\theta)-\psi_{\bot}(r)\sin^2(\theta)],
\label{eq:sig2vz_linear}
\end{equation}
where $\theta$ is the angle between ${\mathbf r}$ and the $z$ axis,
\begin{equation}
\psi_{\|}(r) \equiv \frac{H^2f^2}{2\pi^2}\int_0^{\infty}dk\,P_L(k)\Biggl[\frac{\sin(kr)}{kr}-2\Biggl(\frac{\sin(kr)}{(kr)^3}-\frac{\cos(kr)}{(kr)^2}\Biggr)\Biggr]
\end{equation}
and
\begin{equation}
\psi_{\bot}(r) \equiv \frac{H^2f^2}{2\pi^2}\int_0^{\infty}dk\,P_L(k)\Biggl[\frac{\sin(kr)}{(kr)^3}-\frac{\cos(kr)}{(kr)^2}\Biggr].
\end{equation}
This is a special case of a result derived by
G\'orski (1988).  If the wavenumbers in the integrals are comoving,
as is the convention, the rms error in comoving $z$ separation due
to peculiar velocities is $\sigma_z({\mathbf r})=\sigma_{v_z}({\mathbf r})/H$.
Inserting the $\Gamma=0.2$ linear powerspectrum
of Bardeen et al. (1986) into these equations, normalizing
to $\sigma_8=0.9$ at redshift $z=0$ (i.e., to $\sigma_8\simeq 0.29$
at $z=3$, appropriate for $\Omega_M=0.3$, $\Omega_\Lambda=0.7$),
and integrating numerically, I find the values of $\sigma_z$ shown
in figure~\ref{fig:sigmaz}.  

\subsection{Upshot}
Random peculiar velocities are likely to be the dominant source
of uncertainty in a volume element's distance to the QSO.
The uncertainty in the
element's proper $z$ position is $\sigma_z\sim 1$ proper Mpc,
which corresponds to a time-scale of $t\sim 2\sigma_z/c\sim 7$ Myr.
It might be possible,  
with very high signal-to-noise spectra for a very large number of background sources,
to trace and correct these 
distortions to the time delay surface.  
A number of interesting applications would then be possible.
With current technology, however, we will have accept that
the mean transmissivity in any bin we devise will be
sensitive to the QSO's luminosity over a $\sim 10$ Myr range
of lookback times.  Time resolution significantly better than
this does not appear to be achievable.\footnote{Except at
very early times, $0>t>-20$ Myr, when the positional uncertainties
become increasingly aligned with the time-delay contours; see
figure~\ref{fig:sigmaz}.}

\section{SUMMARY AND DISCUSSION}
\label{sec:summary}
This paper showed that changes over time in the luminosity of a QSO
at redshift $z\sim 3$ 
will produce ionization gradients in the IGM and alter
the Lyman-$\alpha$ forest absorption spectra of background
galaxies in an observable way.  
Because the density of detectable galaxies (${\cal R}\simlt 25$)
at $z\sim 3$ is high, $\sim 1\, {\rm arcmin}^{-2}$, their absorption
spectra can provide a detailed view of the ionization gradients.
If an isotropically radiating QSO has
an AB magnitude at 912\AA\ of $m_{912}=18$, significant decreases
in its luminosity at larger look-back times will be detectable
if they happen $1\simlt t\simlt 50$ Myr before the time of
observation.  The time limits expand for brighter QSOs and shrink
for fainter.  Increases in the QSO's luminosity over this time period
will be harder to
detect than decreases, but since $m_{912}=18$ corresponds to the steep bright end
of the QSO luminosity distribution, they must be much rarer.
\S~\ref{sec:synthesis} sketches out the method and presents
the uncertainties for a simulated QSO with known radiative history $L(t)$;
the section is aimed at those who want more detail but are reluctant 
to read the entire paper. 

The method gives us sensitivity to changes in a QSO's luminosity
over a useful range of times.  Statistical arguments mentioned
in the introduction show that QSOs must change their luminosities
significantly on time-scales $t\simlt 100$ Myr.
If these changes happen on time-scales
$t\simlt 0.1$--1 Myr, a handful of QSOs in large
(SDSS-sized) samples will show major brightness
changes from one decade or century to the next
(Martini \& Schneider 2003).  The method I have described
cannot detect luminosity changes that happen on time-scales
so short (\S~\ref{sec:binsize}), but is sensitive
changes throughout the rest of the allowed range ($1\simlt t\simlt 100$ Myr).
Taken together, the two methods will be able to pin down the
typical QSO lifetime in a robust and direct way.

A number of other authors (e.g., Crotts 1989; Dobrzycki \& Bechtold 1991;
M\/oller \& Kjaergaard 1992;
Fern\'andez-Soto et al. 1995; Liske \& Williger 2001;
Jakobsen et al. 2003; Schirber, Miralda-Escud\'e, \& McDonald 2004;
Croft 2004) have attempted to measure QSO lifetimes
with a similar approach.  Their results were ambiguous.
The reason is that they used the absorption lines in a single background QSO 
(or, in some
cases, a handful) to search for ionization gradients around the foreground
QSO.  When the QSO pair had a small projected separation,
the analyses
were confused by the high densities and large peculiar velocities
near the foreground QSO (see, e.g., figure~\ref{fig:vzeldo}); when
the projected separation was large the foreground QSO's weak effect
on the IGM could not stand out above the cosmic variance.  
This paper's method sidesteps
these difficulties by using numerous faint galaxies rather than a small number
of bright QSOs as the background sources.  As shown
in \S~\ref{sec:uncertainties}, the signal-to-noise ratio of the
background objects' spectra does not affect the final result by much,
but the number of background sources does.  Choosing galaxies as the
background sources is therefore the sensible approach.
With numerous background sources, the weak effect of a QSO on distant
intergalactic matter can be detected with reasonable significance,
the complicated region closest to the QSO can be ignored altogether,
and the peculiar velocity and density gradients at slightly larger
distances can
be compensated by comparing to the amount of Lyman-$\alpha$ absorption
in the large ``control region'' that is illuminated by light emitted by
the QSO at $t\sim 0$ (\S~\ref{sec:delay}).  
The approach I have described should therefore be a significant
improvement over previous work.

Observers who would like to apply this approach in practice should
be aware that the optimal proper size of the observed region 
can be very different from the naive guess
$ct_{\rm max}$ with $t_{\rm max}$ the
maximum time delay of interest.  As $t_{\rm max}\to 0$
intergalactic absorption at radii $R\gg ct_{\rm max}$ becomes
increasingly important for the analysis, as figures~\ref{fig:qpedelay}
and~\ref{fig:recovered_flux.smallr} show.  Two results
derived in \S~\ref{sec:uncertainties} are also relevant.
They are discussed more fully in \S~\ref{sec:commentary}, but
may be summarized as follows:
(A) Galaxy spectra with $S$:$N\sim 4$ per 7\AA\ bin
will be sufficient for measuring the QSO's radiative history with
the stated precision.  Obtaining better spectra will not improve the result
by much.  (B) Cosmic variance places a fundamental limit on the
accuracy of the method.  Only a small part of the universe is bathed in
the light that the QSO emitted at a particular moment in its history,
and the HI content of this region will stray from the global average
for reasons that have nothing to do with the luminosity of the QSO.
The only detectable luminosity changes are those large enough to
alter the intergalactic HI
content by more than its intrinsic random fluctuations $\sigma_V$.
A corollary is that the number of background sources does not
need to be incredibly high; one needs only enough to measure
the mean transmissivity of a region with a precision similar
to $\sigma_V$, and this is easily achieved with spectra of a few dozen
galaxies (but see below).  

I have neglected an important complication in the discussion so far.
It is possible that the QSO's ionizing radiation will not
emerge isotropically but will instead be focused into a bipolar beam with
opening angle $\alpha\sim 90^{\rm o}$ (Barthel 1989).
If the beam were pointed towards earth there would be almost
no effect on the analysis
(see figure~\ref{fig:qpedelay_rad}), but in the typical case
the intergalactic volume that is affected by the QSO's radiation
will be only $\sim 30$\% as large as I have assumed.  
This will increase the uncertainty in the results.  Shrinking
the radius of one of our idealized cylindrical bins (\S\S~\ref{sec:uncertainties}
and~\ref{sec:synthesis}) until it contains only
$30$\% of its previous volume will increase its uncertainty
due to cosmic variance, $\sigma_V$, by a factor of $\sim 1.5$.
If cosmic variance is still the dominant source of noise in the smaller bins,
the error bars in the $\bar f$ vs. $t$ curve will increase
by a similar factor.  Fortunately this is not enough to prevent
us from detecting changes in the QSO's luminosity for $-50{\rm Myr}\simlt t$
(figure~\ref{fig:recovered_flux}).  Moreover, even with the enlarged
error bars it should be easy to distinguish intergalactic volumes
that are illuminated by the QSO's ionizing radiation from neighboring
regions with transmissivities close to the global mean of $\bar f\sim 0.67$,
at least at small radii where the QSO's radiation is most intense
(see, e.g., figure~\ref{fig:recovered_flux.smallr}, which shows that
the QSO's influence
at small radii should be detected with high significance).
This statement is independent of the QSO's radiative history as long
as it has been shining for more than $\sim 10$ Myr, since so large
a volume is (potentially) illuminated by light emitted
at $-10{\rm Myr}<t<0$.  If one were still worried about the possibility
of beaming, QSOs could be chosen for study only if they
have characteristics that suggest their beam is likely to be pointed
towards the earth (e.g., BAL QSOs or radio-loud QSOs).  I suspect that
it will be more interesting to observe many types of QSOs and use
this approach as a direct test of unified models of AGN.  

In any case,
the possibility of beaming makes it clear that the ideal number of
background sources is in fact several times larger than the arguments
of point (B) suggest (see above).  We would like to be cosmic-variance
limited in even the possibly small fraction of the field that is
struck by the QSO's ionizing radiation.  Several hundred background sources
would be ideal.  Fortunately multi-object spectrographs with the
required large field-of-view (e.g., IMACS; Dressler et al. 2003)
can obtain this many spectra with a small number of slitmasks.  Achieving
the necessary signal-to-noise ratio for the background galaxies
at $z\simgt 3$ is less daunting than it sounds, since the large field-of-view
allows one to pick sources from the bright end of the luminosity distribution.

I should mention in closing that
the results of this paper could be extended or improved in a number of ways.
The expected cosmic variance $\sigma_V$ played a large role in the analysis,
yet was estimated from an imperfect model of the three-dimensional
transmissivity power-spectrum.  Better numbers should be derived
from numerical simulations.  It would be interesting to know
if measuring the radiation intensity with something
other than the mean transmissivity $\bar f$ would let us achieve
tighter constraints on the QSO's radiative history before we were limited
by cosmic variance.  I assumed that the QSO would have
redshift $z=3.0$, but in fact observers could choose to observe
QSOs at any redshift $2\simlt z\simlt 4$ where the Lyman-$\alpha$ forest
is visible from the ground and reasonable numbers of background galaxies
can be identified.  Finding the QSO redshift that minimizes the required
observing time would allow one to study a larger number of targets.
This work is admittedly unfinished.  I hope to have shown that it is
worth pursuing further. 

\bigskip
\bigskip
George Becker and Luis Ho were a useful sources of information about AGN
and the uncertainty
in different QSO-redshift estimators.  Esther Hu helped me find
electronic version of her Voigt-profile line lists.
Chuck Steidel fielded my random questions like a pro.
Alice Shapley's email responses to my questions were always
prompt and enlightening, and in addition she graciously allowed me to show the
data in Figure~\ref{fig:threedeepgals}.
Rob Simcoe had helpful advice on several subjects.
Paul Martini offered useful comments on an earlier draft.
Andrew Carnegie's generosity put the food on my table.
It is a pleasure to acknowledge the hospitality of Las Campanas Observatory
during a long stay in Dec 2003 when this paper was begun.

\end{document}